\documentclass[twocolumn]{aastex63}
\received{May 14, 2021}
\revised{}
\accepted{}
\submitjournal{ApJ}

\shorttitle{Characterizing the morphologies of molecular clouds in the Milky Way}
\shortauthors{Yuan et al.}
\begin{document}

\title{A Morphological Classification of 18190 Molecular Clouds
 Identified in $^{12}$CO Data \\ from the MWISP Survey}

\correspondingauthor{Ji Yang}
\email{jiyang@pmo.ac.cn}

\author[0000-0002-0786-7307]{Lixia Yuan}
\email{lxyuan@pmo.ac.cn}
\affiliation{Purple Mountain Observatory and Key Laboratory of Radio Astronomy, Chinese Academy of Sciences, \\
10 Yuanhua Road, Qixia District, Nanjing 210033, PR China}

\author{Ji Yang}
\affiliation{Purple Mountain Observatory and Key Laboratory of Radio Astronomy, Chinese Academy of Sciences, \\
10 Yuanhua Road, Qixia District, Nanjing 210033, PR China}

\author{Fujun Du}
\affiliation{Purple Mountain Observatory and Key Laboratory of Radio Astronomy, Chinese Academy of Sciences, \\
10 Yuanhua Road, Qixia District, Nanjing 210033, PR China}

\author{Xunchuan Liu} 
\affiliation{Department of Astronomy, Peking University, 5 Yiheyuan Road, Haidian District, Beijing 100871, PR China} 

\author{Shaobo Zhang}
\affiliation{Purple Mountain Observatory and Key Laboratory of Radio Astronomy, Chinese Academy of Sciences, \\
10 Yuanhua Road, Qixia District, Nanjing 210033, PR China}

\author{Zehao Lin}
\affiliation{Purple Mountain Observatory and Key Laboratory of Radio Astronomy, Chinese Academy of Sciences, \\
10 Yuanhua Road, Qixia District, Nanjing 210033, PR China}

\author{Jingfei Sun}
\affiliation{Purple Mountain Observatory and Key Laboratory of Radio Astronomy, Chinese Academy of Sciences, \\
10 Yuanhua Road, Qixia District, Nanjing 210033, PR China}

\author{Qing-Zeng Yan}
\affiliation{Purple Mountain Observatory and Key Laboratory of Radio Astronomy, Chinese Academy of Sciences, \\
10 Yuanhua Road, Qixia District, Nanjing 210033, PR China}

\author{Yuehui Ma}
\affiliation{Purple Mountain Observatory and Key Laboratory of Radio Astronomy, Chinese Academy of Sciences, \\
10 Yuanhua Road, Qixia District, Nanjing 210033, PR China}

\author{Yang Su}
\affiliation{Purple Mountain Observatory and Key Laboratory of Radio Astronomy, Chinese Academy of Sciences, \\
10 Yuanhua Road, Qixia District, Nanjing 210033, PR China}

\author{Yan Sun}
\affiliation{Purple Mountain Observatory and Key Laboratory of Radio Astronomy, Chinese Academy of Sciences, \\
10 Yuanhua Road, Qixia District, Nanjing 210033, PR China}

\author{Xin Zhou}
\affiliation{Purple Mountain Observatory and Key Laboratory of Radio Astronomy, Chinese Academy of Sciences, \\
10 Yuanhua Road, Qixia District, Nanjing 210033, PR China}



\begin{abstract}
We attempt to visually classify the morphologies of 18190 molecular clouds, which are identified in the $^{12}$CO(1-0) spectral line data over $\sim$ 450 deg$^{2}$ of the second Galactic quadrant from the Milky Way Imaging Scroll Painting project (MWISP). Using the velocity-integrated intensity maps of the $^{12}$CO(1-0) emission, molecular clouds are first divided into unresolved and resolved ones. The resolved clouds are further classified as non-filaments or filaments. Among the 18190 molecular clouds, $\sim$ 25 $\%$ are unresolved, $\sim$ 64$\%$ are non-filaments, and $\sim$ 11$\%$ are filaments. In the terms of the integrated flux of $^{12}$CO(1-0) spectra of the whole 18190 molecular clouds, $\sim$ 90$\%$ are from filaments, $\sim$ 9$\%$ are from non-filaments, and the rest $\sim$ 1$\%$ are from unresolved sources. Although non-filaments are dominant in the number of the discrete molecular clouds, filaments are the main contributor of $^{12}$CO emission flux. We also present the number distributions of physical parameters of the molecular clouds in our catalog, including their angular sizes, velocity spans, peak intensities of $^{12}$CO(1-0) emission, and $^{12}$CO(1-0) total fluxes. We find that there is a systematic difference between the angular sizes of the non-filaments and filaments, with the filaments tending to have larger angular scales. The H$_{2}$ column densities of them are not significantly different. We also discuss the observational effects, such as those induced by the finite spatial resolution, beam dilution and line-of-sight projection, on the morphological classification of molecular clouds in our sample. 
\end{abstract}



\section{Introduction} \label{sec:intro}
Molecular clouds usually exhibit complex and hierarchical structures, whose boundaries can be defined by their CO emission at the lower rotational transitions \citep{Heyer2015}. The spatial power spectrum \citep{Ingalls2004, Gazol2010} and the $\Delta$-variance \citep{Stutzki1998, Bensch2001, Elia2014, Dib2020} techniques have been used to quantitatively analyze the fractal structures of molecular clouds.
 \citet{Falgarone1991, Stutzki1998} have characterized the boundaries of clouds using fractal dimensions and gave a value of about 1.5 for the two-dimensional (2-D) projections, and $\sim$ 2 in the three-dimensional (3-D) structures \citep{Beattie2019}. 
\cite{Dib2020} investigated the structures of Cygnus-X North molecular cloud using the $\Delta$-variance spectrum and indicated its characteristic scales were in a range of $\sim$ 0.5 -- 1.2 pc.

Since the systematic discovery of filamentary structures in molecular clouds by the Herschel observations \citep{Molinari2010, Andre2010}, more and more attentions have been focused on filaments. Numerous studies have revealed that star-forming clouds usually exhibit filamentary structures, within which a significant fraction of gravitationally bound dense cores and protostars are embedded \citep{Andre2013, Andre2014, Andre2017, Schneider2012, Contreras2016, Yuan2019, Yuan2020}. There are also a fair amount of non-filaments, e.g., globules, cometary, and extended structures \citep{Bok1947, Bourke11995, Makela2013, Goicoechea2020}, which either harbor a few young stars or are less opaque, less dense, with little or no star formation \citep{Bourke11995, Bourke21995, Reipurth2008, Haikala2010, Goicoechea2020}. However, the percentages and mass fractions of filaments and non-filaments in molecular clouds are still unclear. It is interesting and important to explore the morphologies of molecular clouds, e.g., to investigate the fraction of filaments and non-filaments in the molecular cloud populations.

A large-scale, unbiased, and high-sensitive spectral survey is essential to investigate the morphologies of molecular clouds in a wide range of the spatial scales. The ongoing unbiased Galactic plane CO survey, the Milky Way Imaging Scroll Painting (MWISP), provides us opportunities to promote the research of the morphologies of molecular clouds to a huge sample with rich details. The mapping area of MWISP covers the Galactic longitude from $l =$ 9.75$^{\circ}$ to 230.25$^{\circ}$ and the Galactic latitude from $b =$ -5.25$^{\circ}$ to 5.25$^{\circ}$. This CO survey is performed using the 13.7 m telescope of Purple Mountain Observatory and observes $^{12}$CO, $^{13}$CO, and C$^{18}$O (J= 1-0) lines simultaneously \citep{Su2019}.

In this paper, we focus on the unbiased catalog of molecular clouds in the Galactic plane with 104.75$^{\circ}$ $< l <$ 150.25$^{\circ}$, $-$5.25$^{\circ}$ $< b <$ 5.25$^{\circ}$, and $-$95 km s$^{-1}$ $< V_{\rm LSR} <$ 25 km s$^{-1}$ provided by the MWISP \citep{Yan2021}. This catalog consists of 18190 molecular clouds. We attempt to perform a morphological classification to this sample through visual inspection. These molecular clouds are first divided into the resolved and unresolved ones, the resolved clouds are further classified as non-filaments and filaments. This paper is organized as follows. Section 2 introduces the $^{12}$CO(1-0) spectral line data and the unbiased catalog of molecular clouds. Section 3 presents the morphological classification for molecular clouds, including the criteria, processes and results of classification. In section 4, we compare the physical properties of filaments and non-filaments. Section 5 discusses the effects of the finite spatial resolution, beam dilution and 2-D projection on the morphological fractions of molecular clouds.

\begin{figure*}
\plotone{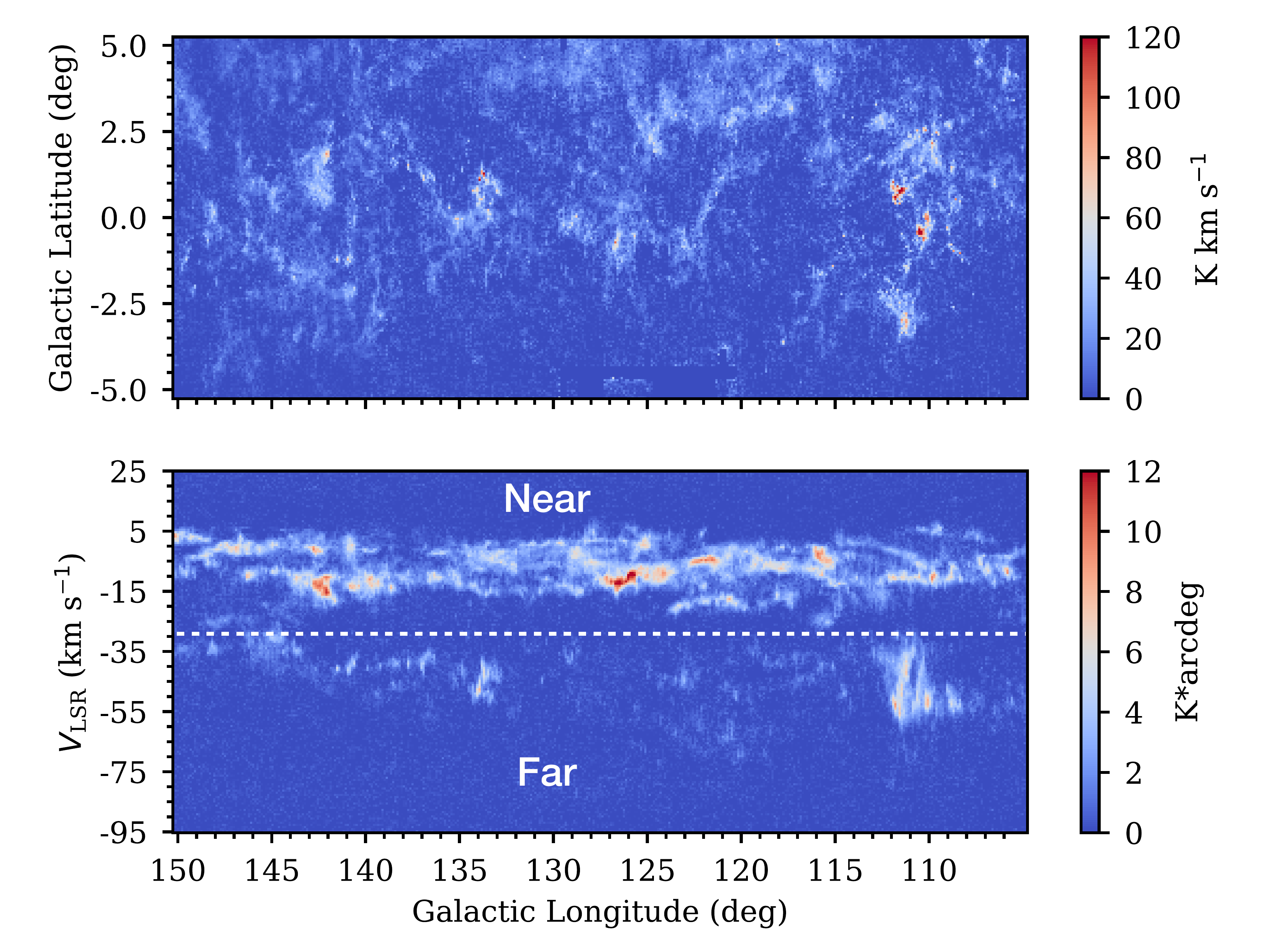}
\caption{\textbf{Top panel}: the velocity-integrated intensity map of $^{12}$CO(1-0) emission in the second Galactic quadrant with 104.75$^{\circ}$ $< l <$ 150.25$^{\circ}$ and $|b| <$ 5.25$^{\circ}$, the integrated velocity ($V_{\rm LSR}$) ranges from $-$95 km s$^{-1}$ to 25 km s$^{-1}$. 
\textbf{Bottom panel}: the latitude-integrated intensity map of $^{12}$CO(1-0) emission in the region with 104.75$^{\circ}$ $< l <$ 150.25$^{\circ}$ and $-$95 km s$^{-1}$ $< V_{\rm LSR} <$ 25 km s$^{-1}$, the integrated latitude ranges from $-$5.25$^{\circ}$ to 5.25$^{\circ}$. The white-dashed line at $V_{\rm LSR} =$ -30 km s$^{-1}$ seperates the molecular clouds into two groups, i.e., Near and Far. } 
\label{fig:fco}
\end{figure*}

\begin{figure*}
\plotone{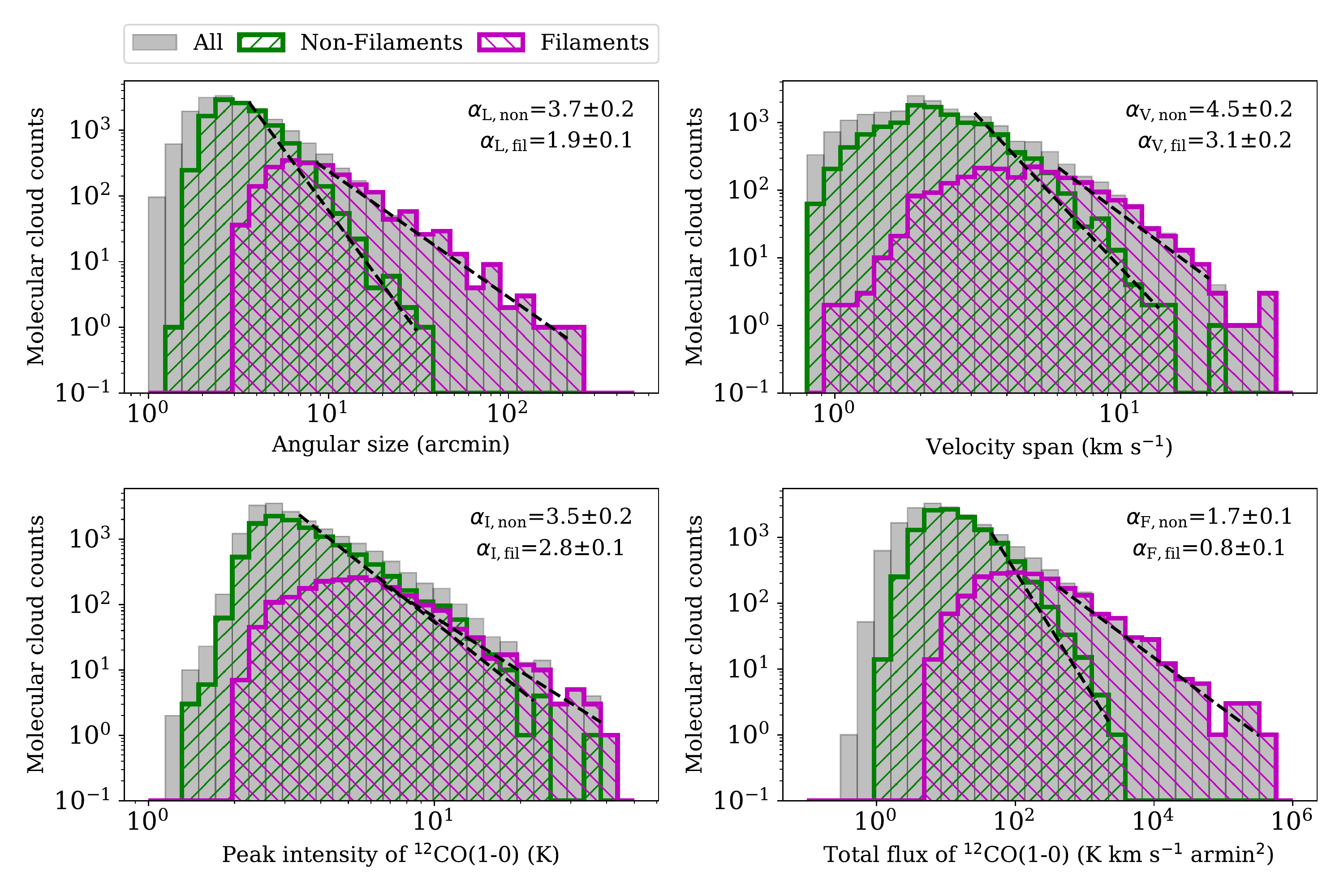}
\caption{The number distributions of angular size, velocity span, the peak intensity of $^{12}$CO emission, and the total flux of $^{12}$CO emission for molecular clouds. The gray bars indicate the whole 18190 molecular clouds in our catalog, including unresolved sources, non-filaments, and filaments. The green histograms represent molecular clouds showing non-filaments, and the magenta ones are for filaments. The dashed-black lines fit the power-law distributions, dN/dX $\propto$ X$^{-(\alpha_{\rm X}+1)}$, for filaments and non-filaments, whose values of $\alpha_{\rm X}$ are noted in each panel.}
\label{fig:fpara}
\end{figure*}

\section{Data}
\subsection{The $^{12}$CO(1-0) emission in the second Galactic quadrant $(104.75^{\circ} < l < 150.25^{\circ}$, $|b| < 5.25^{\circ}$, and $-95$ $km$ $s^{-1}$  $< V_{\rm LSR} < 25$ $km$ $s^{-1})$}

The large-scale distribution of the molecular gas across the second Galactic quadrant of 450 deg$^{2}$ is showed in Figure \ref{fig:fco}. The upper colormap represents the velocity-integrated intensity of $^{12}$CO(1-0) line emission covering 104.75$^{\circ}$ $< l <$ 150.25$^{\circ}$, $|b| <$ 5.25$^{\circ}$. The integrated velocity range is $V_{\rm LSR}$ from $-$95 km s$^{-1}$ to 25 km s$^{-1}$. The lower colormap shows the distribution of the latitude-integrated intensity of $^{12}$CO(1-0) line. The integrated latitude range is $b$ from $-$5.25$^{\circ}$ to 5.25$^{\circ}$. To display the gas structures clearly, we divided the whole region into three parts, whose Galactic longitudes are in the range of $l =$ (104.75$^{\circ}$, 120$^{\circ}$), (120$^{\circ}$, 135$^{\circ}$), and (135$^{\circ}$, 150.25$^{\circ}$), respectively. Their velocity-integrated intensity and latitude-integrated intensity maps are presented in Figure \ref{fig:fco_partI} -- \ref{fig:fco_partIII}, respectively. The molecular gas distributes inhomogeneously and hierarchically.
In the Longitude-Velocity diagram, molecular gas mainly concentrates in the range of $V_{\rm LSR}$ $=$ ($-$25, 10) km s$^{-1}$ and ($-$ 65, $-$ 30) km s$^{-1}$. Also, the well-known molecular clouds located within the map-regions have been noted. Parts of the observations in this region previously were published in \citet{Yan2021, Sun2020, Du2016, Du2017}.

\begin{deluxetable*}{lcrrccccl}
\tablenum{1}
\tablecaption{A catalog of molecular clouds \label{tab:t1}}
\tablewidth{0pt}
\tablehead{
\colhead{Name} & \colhead{$l_{\rm cen}$} & \colhead{$b_{\rm cen}$} & \colhead{$V_{\rm LSR}$} & \colhead{$\mathcal{A}$} & \colhead{$V_{\rm span}$} &\colhead{Peak Intensity} & \colhead{Flux} & \colhead {Morphology} \\
\colhead{Number} & \colhead{(degree)} & \colhead{(degree)} & \colhead{(km s$^{-1}$)} &
 \colhead{(arcmin$^{2}$)} &\colhead{km s$^{-1}$} &\colhead{(K)} & \colhead{(K km s$^{-1}$ arcmin$^{2}$)} & \colhead{Type}
}
\decimalcolnumbers
\startdata
G105.434+00.481-050.97 & 105.434 & 00.481 & -50.97 & 02.75 & 01.50 & 2.4 & 1.73 & Unresolved \\
G105.435+00.996-012.20 & 105.435 & 00.996 & -12.20 & 07.75 & 02.00  & 3.8 & 10.38 & Non-filament \\
G105.441+03.708-008.23 & 105.441 & 03.708 & -8.23 & 04.00 & 01.67  & 2.8 & 4.15 & Unresolved \\
G105.445+03.776-047.29 & 105.445 & 03.776 & -47.29 & 29.25 & 03.00 &10.2 & 103.38 & Non-filament \\
G105.445+00.338-042.96 & 105.445 & 00.338 & -42.96 & 05.25 & 03.51 & 3.2 & 10.51 & Non-filament \\
G105.449+00.579-042.87 & 105.449 & 00.579 & -42.87 & 20.25 & 04.17 & 5.3 & 53.46 & Non-filament \\
G105.459+00.514-045.21 & 105.459 & 00.514 & -45.21 & 04.50 & 02.34 & 2.3 & 5.04 & Non-filament \\
G105.461+00.062-011.38 & 105.461 & 00.062 & -11.38 & 22.75 & 02.51 & 7.7 & 48.35 & Filament \\
G105.462-02.225-025.45 & 105.462  & -02.225  & -25.45 & 02.00 & 01.17 & 2.8 & 2.08 & Unresolved \\
G105.466+03.014-068.21 & 105.466 & 03.014 & -68.21 & 05.75 & 03.51 & 2.9 & 9.19 & Non-filament \\
G105.466+00.593-048.35 & 105.466 & 00.593 & -48.35 & 103.25 & 04.34 & 6.8 & 320.12 & Filament \\
\enddata
\tablecomments{The central Galactic coordinates ($l_{\rm cen}$, $b_{\rm cen}$) for each cloud are the averaged Galactic coordinates in its velocity-integrated $^{12}$CO(1-0) intensity map, weighting by the value of the velocity-integrated $^{12}$CO(1-0) intensity. 
The central velocity ($V_{\rm LSR}$) for each cloud is the averaged radial velocity in its radial velocity field, weighting by the value of the velocity-integrated $^{12}$CO(1-0) intensity. The $\mathcal{A}$ is the angular area of each cloud, which is projected along the line of sight. The V$_{\rm span}$ represents the velocity span of each cloud cube in the velocity axis of PPV space, which is calculated using the number of velocity channels in the cloud cube multiplied by a velocity resolution of 0.158 km s$^{-1}$.
Peak Intensity is the peak value of $^{12}$CO(1-0) line intensity in each cloud.
The Flux is the integrated flux of $^{12}$CO(1-0) line emission for each cloud, Flux= $\int$T$_{\rm mb}(l,b,v)dldbdv$ = 0.167$\times$0.25$\Sigma$T$_{\rm mb}(l,b,v)$ K km s$^{-1}$ arcmin$^{2}$, where T$_{\rm mb}(l, b, v)$ is the $^{12}$CO line intensity at the coordinate of $(l, b, v)$ in PPV space, $dv =$ 0.167 km s$^{-1}$ is the velocity resolution, $dldb =$ 0.5 arcmin $\times$ 0.5 arcmin $=$ 0.25 arcmin$^{2}$, the angular size of a pixel is 0.5 arcmin. This table is available in its entirety from the online journal. A portion is shown here for guidance regarding its form and content.}
\end{deluxetable*}

\subsection{An unbiased catalog of molecular clouds}
In this work, a molecular cloud is defined as a connected structure in a 3-D space. In practice, a set of contiguous voxels in the position-position-velocity (PPV) cube with the intensities of $^{12}$CO emission above a certain threshold are extracted as a molecular cloud. The extracting method we used was developed by \cite{Yan2020}, based on the Density-Based Spatial Clustering of Applications with Noise (DBSCAN) algorithm, which was originally designed to discover clusters of arbitrary shape \citep{ester1996}.

\cite{Yan2021} has completed the identification of molecular clouds using the DBSCAN algorithm in the second Galactic quadrant with 104.75$^{\circ}$ $< l <$ 150.25$^{\circ}$, $|b| <$ 5.25$^{\circ}$, and $-95$ km s$^{-1}$  $<$ $V_{\rm LSR}$ $<$ 25 km s$^{-1}$. We check the averaged $^{12}$CO(1-0) line profile and the velocity-integrated intensity map for each cloud, 18190 out of 18503 molecular clouds can be used for further analysis. Among the 313 excluded structures, 88 are fake structures caused by bad channels, which are mainly located at the velocity of $-$24 km s$^{-1}$, and 225 clouds located at the edges of the region ($l =$ 104.75$^{\circ}$ or 150.25$^{\circ}$, $b = \pm$ 5.25$^{\circ}$, $V_{\rm LSR} = $25 or - 95 km s$^{-1}$) are incomplete in this work.

The basic parameters for each molecular cloud are listed in Table \ref{tab:t1}, including the central Galactic coordinate, the peak intensity and flux of $^{12}$CO line emission, the projected angular area ($\mathcal{A}$), and its velocity span ($V_{\rm span}$, which is the velocity span of molecular cloud cube in the velocity axis of PPV space).
Figure \ref{fig:fpara} presents the number distributions of these parameters. The angular sizes of these molecular clouds, which are calculated as $\sqrt{\mathcal{A}}$, range from 1 arcmin to 246 arcmin and concentrate in the scale of 2.2 -- 4.3 arcmin. Their velocity spans are distributed in (0.5, 43.4) km s$^{-1}$ and mainly located in (1.5, 3.0) km s$^{-1}$.   
The peak intensities of $^{12}$CO(1-0) emission for these molecular clouds are distributed in a wide range of 1.2 -- 50.8 K, although the values for most sources are located in a narrow range of 2.6 -- 4.2 K, their $^{12}$CO(1-0) emission flux distribute in a range of 0.53 -- 3.9$\times$10$^{5}$ K km s$^{-1}$ arcmin$^{2}$ and concentrate in (4.4, 27.4) K km s$^{-1}$ arcmin$^{2}$. The typical range of each parameter, which is from the quantile of 0.25 to 0.75 in its sequential data, has been listed in Table \ref{tab:t2}. Although \cite{Yan2021} has calculated the distance of 76 molecular clouds in this catalog, most of the molecular clouds in this catalog do not have the available distance information. Thus, we group the 18190 molecular clouds into two parts, corresponding to the near and far ranges, respectively. As shown in Figure \ref{fig:fco}, molecular clouds in the near range are in the velocity range of ($-$30, 25) km s$^{-1}$, clouds in the far range are in ($-$95, $-$30) km s$^{-1}$.

\begin{figure*}
\plotone{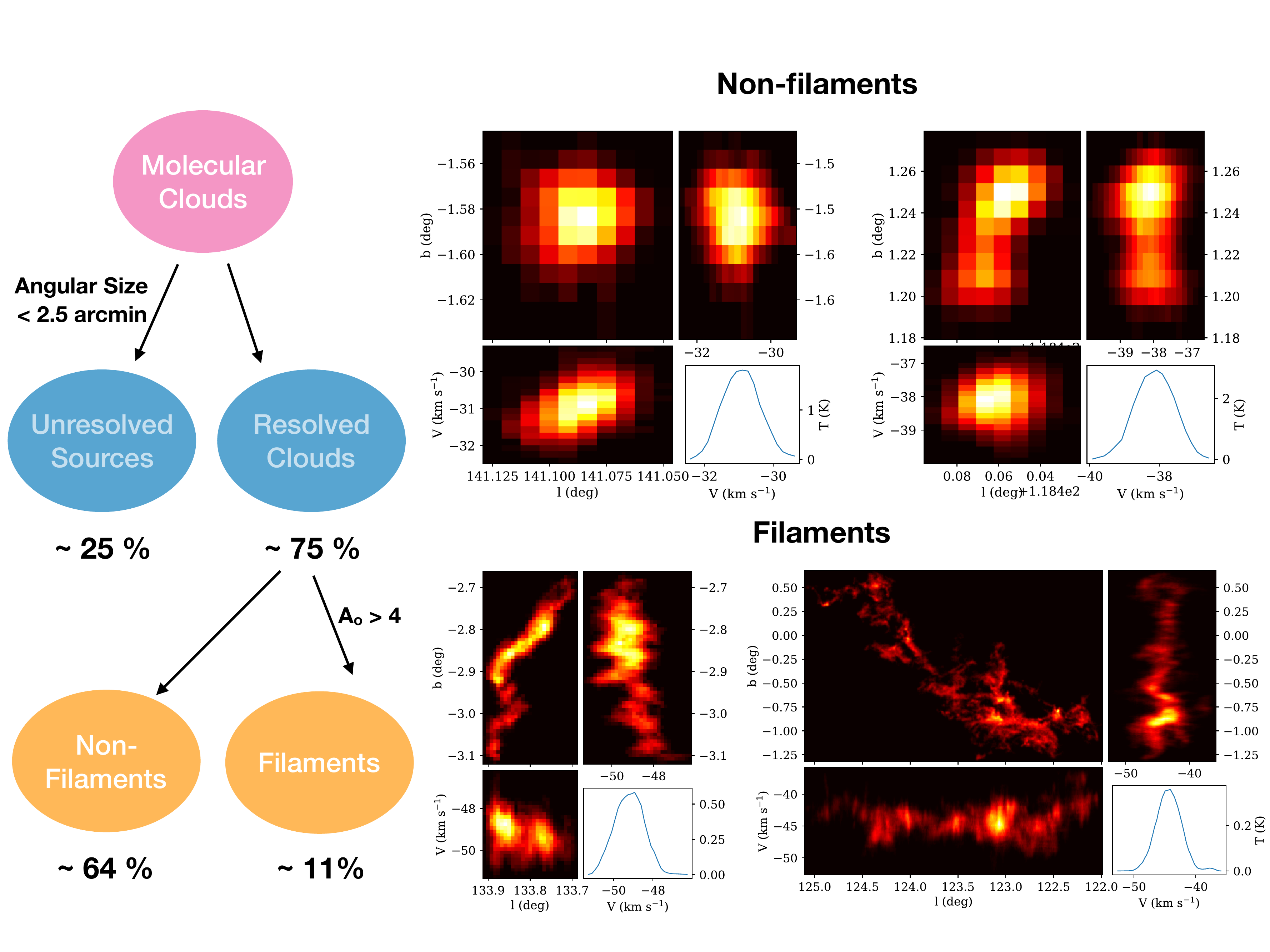}
\caption{A schematic plot of morphological classification for molecular clouds in our catalog.}
\label{fig:fclassifi}
\end{figure*}

\begin{deluxetable*}{lcccc}
\tablenum{2}
\tablecaption{Typical ranges of the physical parameters for molecular clouds in the catalog \label{tab:t2}}
\tablewidth{0pt}
\tablehead{
\colhead{Types}  & \colhead{Angular size} & \colhead{V$_{\rm span}$} &\colhead{Peak Intensity} & \colhead{Flux} \\
& \colhead{(arcmin)} &\colhead{km s$^{-1}$} &\colhead{(K)} & \colhead{(K km s$^{-1}$ arcmin$^{2}$)} 
}
%\decimalcolnumbers
\startdata
Molecular clouds & 2.2 -- 4.3 & 1.5 -- 3.0 & 2.6 -- 4.2 & 4.4 -- 27.3 \\
Non-filaments & 2.6 -- 4.2 & 1.7 -- 3.0 & 2.7 -- 4.2 & 6.7 -- 25.5 \\
Filaments & 5.8 -- 12.1 & 3.2 -- 6.5 & 4.0 -- 7.2 & 50.4 -- 420.1 \\
\enddata
\tablecomments{The typical range for each parameter is from the quantile of 0.25 to 0.75 in its sequential data.}
\end{deluxetable*}

\begin{figure*}
\plotone{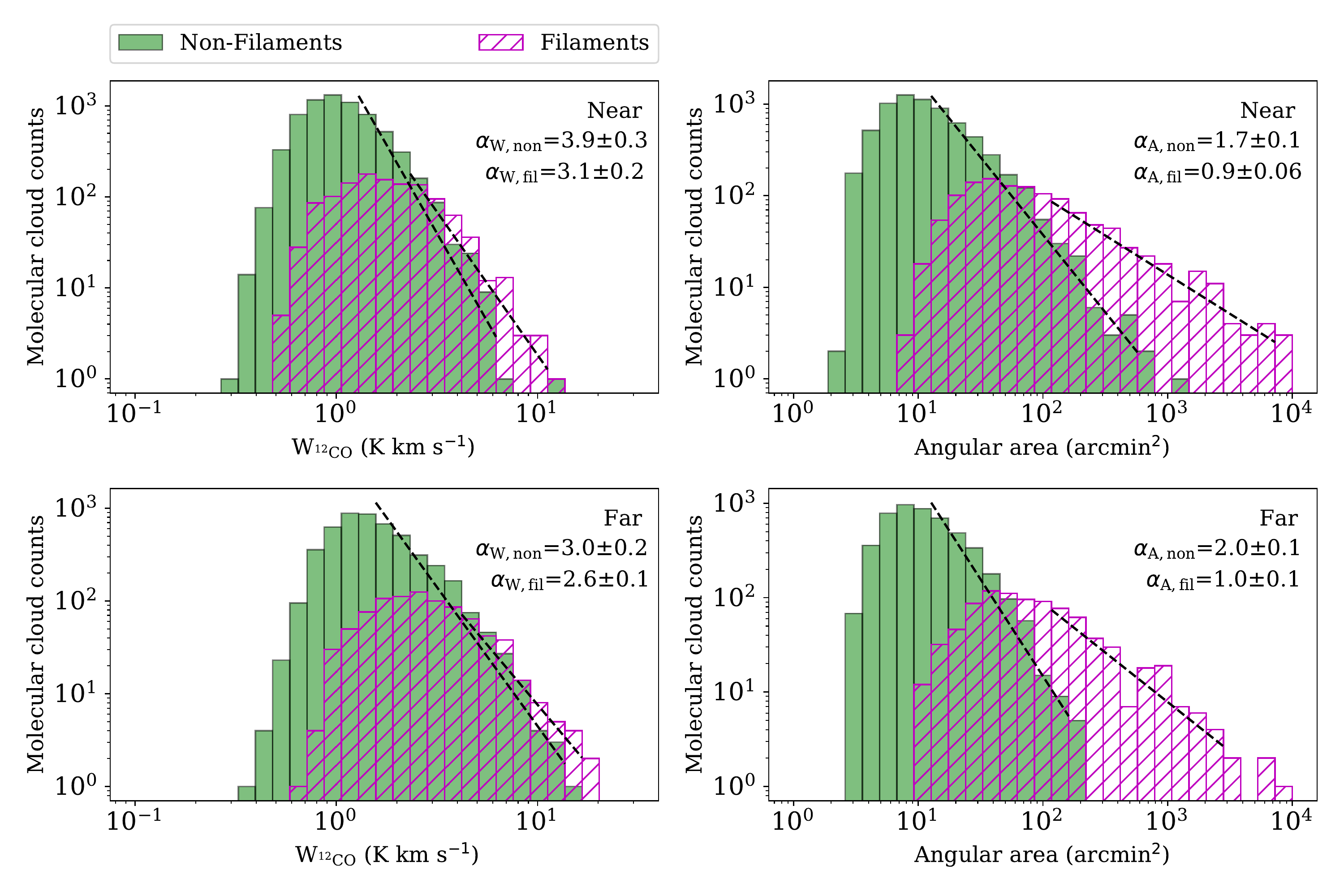}
\caption{\textbf{Upper panel}: the number distributions of the averaged $^{12}$CO velocity-integrated intensity (W$_{\rm ^{12}CO}$) and the angular areas ($\mathcal{A}$) for molecular clouds in the near range. \textbf{Lower panel}: same as above, but for molecular clouds in the far range. The green histograms represent molecular clouds showing non-filaments, and the magenta ones are for filaments. The dashed-black lines fit their power-law distributions, dN/dX $\propto$ X$^{-(\alpha_{\rm X}+1)}$, whose negative slopes are noted in each panel. \label{fig:fproj}}
\end{figure*}

\begin{deluxetable*}{lcccc}
\tablenum{3}
\tablecaption{Typical ranges of angular area ($\mathcal{A}$) and H$_{2}$ column density (N$_{\rm H_{2}}$) for filaments and non-filaments. \label{tab:t3}}
\tablewidth{0pt}
\tablehead{
\colhead{Types}  & \colhead{A} & \colhead{$\alpha_{\rm A}$} & \colhead{N$_{\rm H_{2}}$ } & \colhead{$\alpha_{\rm W}$} \\
& \colhead{(arcmin$^{2}$)} & &\colhead{10$^{20}$ cm$^{-2}$} & \\
& \colhead{Near, Far} & \colhead{Near, Far} & \colhead{Near, Far} & \colhead{Near, Far}
}

\startdata
Non-filaments & 6.5 -- 18, 6.8 -- 17& 1.7$\pm$0.1, 2.0$\pm$0.1 & 1.6 -- 2.8 , 2.2 -- 4.0  & 3.9$\pm$0.3 , 3.0$\pm$0.2  \\
Filaments & 32.0 -- 147.4 , 35.8 -- 144.8  & 0.9$\pm$0.1, 1.0$\pm$0.1 & 2.4 -- 5.2 , 3.4 -- 7.6 & 3.1$\pm$0.2, 2.6$\pm$0.1 \\
\enddata

\tablecomments{The H$_{2}$ column density (N$_{\rm H_{2}}$) are calculated using the N$_{\rm H_{2}}$ = X$_{\rm CO}$W$_{\rm ^{12}CO}$, where X$_{\rm CO}$=2$\times$10$^{20}$ cm$^{-2}$ (K km s$^{-1}$)$^{-1}$. The $\alpha_{\rm A}$ and $\alpha_{\rm W}$ represent the fitted slopes for the power law distributions of the angular sizes and W$_{\rm ^{12}CO}$, respectively. The Near represents the molecular clouds in the near range ($V_{\rm LSR}$ from $-$30 km s$^{-1}$ to 25 km s$^{-1}$), the Far mean the molecular clouds in the far range ($V_{\rm LSR}$ from $-$95 km s$^{-1}$ to $-$30 km s$^{-1}$. The typical range for each parameter is from the quantile of 0.25 to 0.75 in its sequential data.}
\end{deluxetable*}

\section{Morphological classification}
The morphologies of molecular clouds are usually thought to be complex and hierarchical. As a fundamental character of molecular clouds, the morphology still lacks a systematic description. Since a profusion of filaments in the Galactic molecular clouds have been revealed in the Hi-GAL imaging surveys with Herschel \citep{Andre2010, Molinari2010}, and the role of interstellar filaments in the star formation process has been emphasized \citep{Andre2013, Hacar2013, Kirk2013, Andre2014, Schneider2012, Contreras2016}, the filamentary clouds have gained more and more attention. We wonder if all the molecular clouds are filamentary. The MWISP CO survey provides us opportunities to promote the research of the morphologies of molecular clouds to a huge sample with rich details. As the first step in the morphological classification of molecular clouds, after visually inspecting the whole molecular cloud samples in our catalog, we find the main features of molecular clouds can be simplified and labeled as filaments and non-filaments. Thus we attempt to take a systematic and binary classification for the morphologies of molecular clouds. This simplified classification may provide a new dimension for us to further understand molecular clouds. 

For each molecular cloud in our catalog, we draw its $^{12}$CO(1-0) intensity maps integrated along three different directions ($l, b, v$) and its averaged $^{12}$CO(1-0) spectrum. We show these maps of four molecular clouds in Figure \ref{fig:fclassifi}. The morphological classification for these 18190 molecular clouds are performed by visually inspecting their integrated $^{12}$CO(1-0) intensity maps.

\subsection{Classificatory criteria}

After preliminarily visual inspections of the whole samples, we attempt to classify the whole molecular clouds into unresolved sources, non-filaments, and filaments. While \citet{Andre2013, Andre2014} have revealed that the observed filaments in the nearby star-forming clouds of our Galaxy share common properties, and the characters of larger-scale and dense filaments also have been systematically analyzed in \citet{Ligx2013, Wang2015, Wang2016}, a quantitative definition for filaments is still lacking. In this work, according to the beam size of MWSIP CO data, we empirically define a filament segment as an elongated, narrow, or twisty structure, whose aspect ratio (the ratio between its length and width, $A_{o}$) is larger than 4 ($A_{o}$ $\gtrsim$ 4). The length of a filament segment is defined as the length along its spine, its width is estimated as the radial cuts across the spine. Once the length of an elongated, narrow, or twisty structure  is estimated to be larger than 4 times its width by visual inspection, we label this structure as a filament segment. The criteria of our classifications are described as below:

(1) Molecular clouds are divided into unresolved sources and resolved clouds. The unresolved sources are defined as the structures with the projected angular areas less than 2.5 arcmin $\times$ 2.5 arcmin (5 pixel $\times$ 5 pixel), corresponding to a diameter of $\sim$ 3 beam size.

(2) The resolved clouds are further classified as non-filaments or filaments. Filaments are defined as structures including at least one filament segment.

(3) Non-filaments are the resolved molecular clouds, which do not contain the resolved filament segments.

\subsection{Processes and results of morphological classification}
According to the above criteria of morphological classification, three authors independently classified these 18190 molecular clouds by visual inspection. As the scheme illustrated in Figure \ref{fig:fclassifi}, molecular clouds are first divided into unresolved and resolved clouds. 
According to the results from three investigators, the consistent results from at least two investigators are considered to be the types of molecular clouds, i.e., unresolved or resolved sources. After completed the classification of unresolved and resolved clouds, the resolved ones are further visually classified as non-filaments or filaments. We also use the consistent results from at least two investigators as the morphological types of molecular clouds. Finally, in a total of 18190 molecular clouds, we find that the number of unresolved sources is 4448, which takes a percentage of about 25 $\%$, in the rest $\sim$ 75$\%$, there are $\sim$ 64 $\%$ molecular clouds (11680) classified as non-filaments and $\sim$ 11 $\%$ as filaments (2062). In addition, we compare the agreement fractions between the results labeled by arbitrary two of three individuals, the agreements are in the level of 60 $\%$ -- 80 $\%$.

Generally, the morphologies of filamentary clouds include the single filamentary segment, the dominant filament with branches, the intersection of multiple filaments, and the complex network built by a system of filamentary segments and their sub-branches. A single filament is a single structure, which satisfies the definition for the filament segment. For a single dominant filament with relatively few branches, its branches are along the main filament and have smaller lengths and less $^{12}$CO emission intensities than that of the main chunk, somewhat like the branches of a tree. Moreover, for the intersection of multiple filaments and the network connected by a system of filaments, their hierarchical and filamentary structures are complex, but contain at least one filamentary segment. In Figures 12 and 13, we show the examples of molecular clouds classified as filamentary structures.

The morphologies of non-filament structures primarily include clumpy structures and extended structures. Clumpy structures present one clump or are dominated by one clump. For one clump, its $^{12}$CO emission intensity gradually decreases from the center to the outside, i.e., the distribution of $^{12}$CO(1-0) intensity nearly fits the 2D-Gaussian profile. For a structure dominated by one clump, it may also contain a tail or several fibers but with lengths smaller than the size of the dominant clump. Compared with the clumpy structure, the extended structure is more stretched, its shape is similar to a cylinder or an arc, but its aspect ratio is roughly less than 4 ($A_{o}$ $\lesssim$ 4). Figures 14 and 15 represent the examples of non-filaments, corresponding to the clumpy and extended structures, respectively.

We should note that the morphologies of molecular clouds are affected by the spatial resolution of $^{12}$CO spectral lines. The morphologies of molecular clouds in this work are based on the spatial resolution ($\sim$ 50 arcsec) of 13.7 m millimeter-wavelength telescope of the Purple Mountain Observatory (PMO). It is possible that the morphological characters and fractions may change under the higher spatial resolution data.

\section{The characteristics of filaments and non-filaments}
What are the different characteristics between the non-filaments and filaments clouds? Is there any connection between them? To answer these questions, we further present the characteristics of filaments and non-filaments, including their Galactic distributions, angular sizes, velocity spans, the peak intensities and fluxes of $^{12}$CO line, and the averaged velocity-integrated $^{12}$CO line intensity (W$_{\rm ^{12}CO}$). The number distributions of these parameters are presented in Figure \ref{fig:fpara} and \ref{fig:fproj}, and also fitted with the power-law function, as dN/dX $\propto$ X$^{- (\alpha_{\rm X}+1)}$. 
The typical ranges of these parameters, which are calculated from the quantile of 0.25 to 0.75 in their sequential data, are listed in Table \ref{tab:t2} and \ref{tab:t3}.

\subsection{Galactic distributions}
\begin{figure*}
\plotone{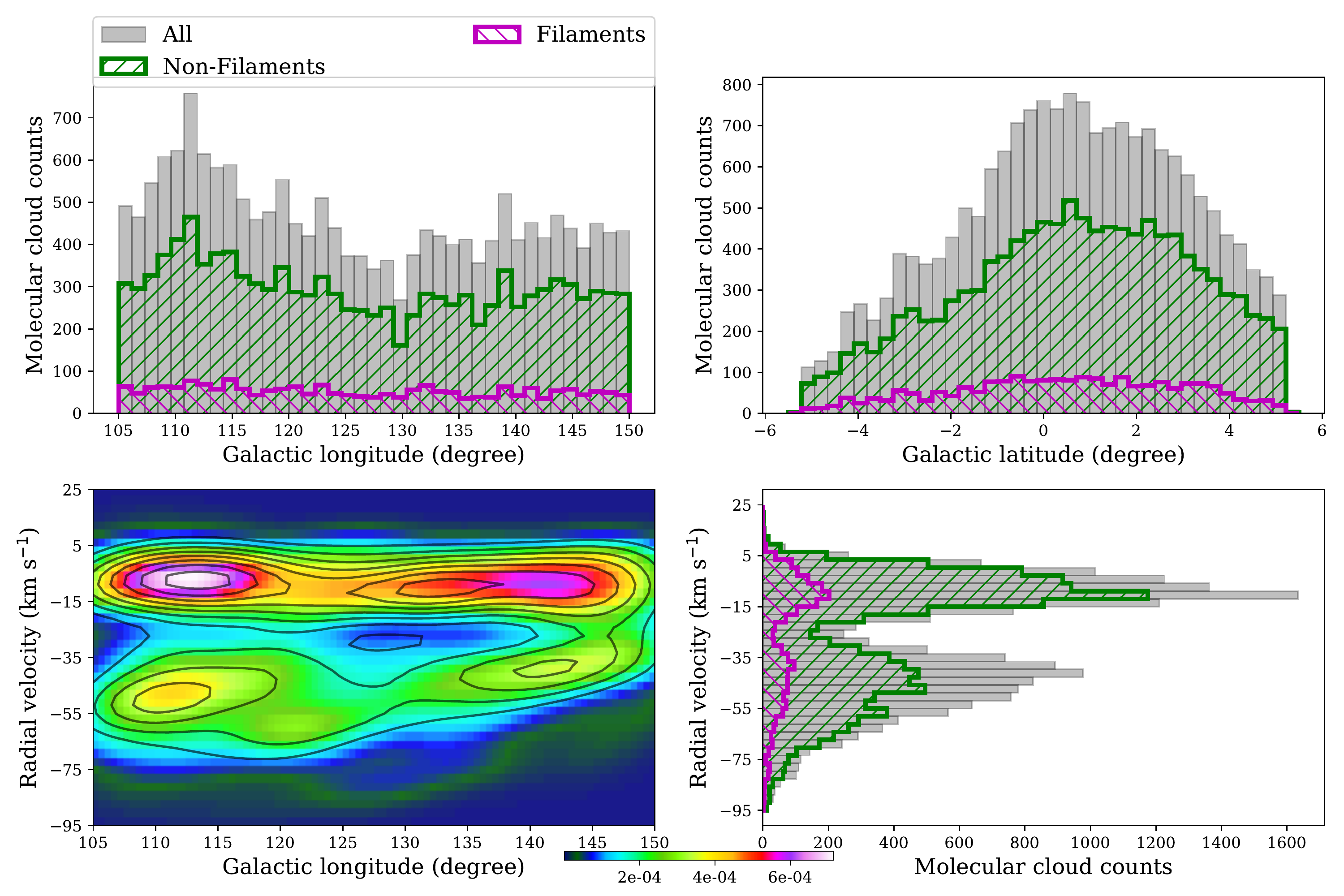}
\caption{\textbf{Upper panel}: the number distributions of molecular clouds in the Galactic longitude and Galactic latitude. \textbf{Lower-left panel}: the colormap represents the 2D-PDF of the number of non-filaments in the longitude-velocity scope, the black contours denote that for filaments, levels are from 25 to 95 percent with an interval of 10 percent of the peak value (5.9$\times$10$^{-4}$). \textbf{Lower-right panel}: the number distribution of molecular clouds in the radial velocity. The grey bars indicate the whole molecular clouds in our catalog, including unresolved sources, non-filaments, and filaments. The green histograms represent molecular clouds showing non-filaments, and the magenta ones are for filaments.}
\label{fig:fpos}
\end{figure*}

Figure \ref{fig:fpos} presents the number distributions of molecular clouds in the Galactic longitude, Galactic latitude, and radial velocity, respectively. Furthermore, we utilize the Kernel-density estimation using Gaussian kernels in the PYTHON package \href{https://docs.scipy.org/doc/scipy/reference/generated/scipy.stats.gaussian\_kde.html}{scipy.stats.gaussian\_kde}, which is a way to estimate the probability density function of a random variable in a non-parametric way (2D-PDF), to demonstrate the 2D-PDF of the number of filaments and non-filaments in the longitude-velocity scope. The resultant 2D-PDF maps are shown in Figure \ref{fig:fpos}.

\begin{figure*}
\plotone{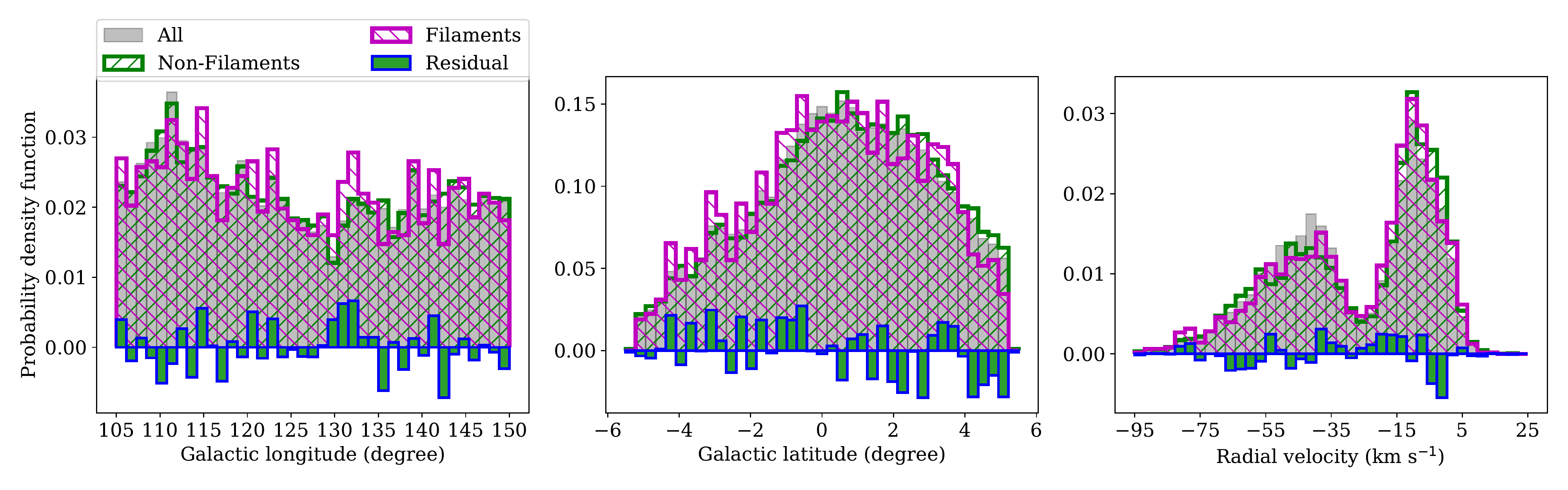}
\caption{The normalized number distributions of molecular clouds in the Galactic longitude, Galactic latitude and radial velocity. The grey bars indicate the whole molecular clouds in our catalog, including unresolved sources, non-filaments, and filaments. The green histograms represent molecular clouds showing non-filaments, and the magenta ones are for filaments. The blue bars indicate the residual of the probability density at the bins between filaments and non-filaments.}
\label{fig:fpos2}
\end{figure*}

To quantitatively compare the Galactic distributions of filaments and non-filaments, we normalized their histograms and presented their probability density function of the numbers in the Galactic longitude, Galactic latitude, and radial velocity in Figure \ref{fig:fpos2}. In the Galactic longitude $l$ from 104.75$^{\circ}$ to 150.25$^{\circ}$, we also perform the Kolmogorov-Smirnov test to compare the number distributions of filaments and non-filaments using the \href{https://docs.scipy.org/doc/scipy/reference/generated/scipy.stats.ks\_2samp.html}{scipy.stats.ks\_2samp} procedure in the scipy package, resulting in a p-value of 0.45. This suggests we can not reject the identical distribution for the number distributions of filaments and non-filaments at 45$\%$ level. In the Galactic latitude $b$ from $-5^{\circ}$ to 5$^{\circ}$, both filaments and non-filaments appear to peak at more positive latitudes, concentrating in a range of ($-1^{\circ}$, 3$^{\circ}$).
This may be attributed to the gas layer in the outer Galaxy is systematically warped away from flatness. The first detection and descriptions of the warped HI gas layer of the Outer Galaxy were given by \citet{Westerhout1957, Burke1957, Kerr1957, Oort1958}, more detailed and quantitative parameters of the HI warp and flares were provided in \citet{Burton1986, Diplas1991, Voskes2006}. Furthermore, \cite{Wouterloot1990} derived the distribution of molecular clouds in the outer Galaxy and found it shows the same warped shapes and flaring thickness as that of the HI gas layer. We also find that the number distribution in the radial velocity for non-filaments is close to that for filaments. Both of them concentrate on the velocity range of (-20, 5) km s$^{-1}$ and (-65, -30) km s$^{-1}$.

\subsection{Angular size}

In Figure \ref{fig:fpara}, for the non-filaments, their angular sizes (L) are in a range of 2 -- 40 arcmin and concentrate on a range of $\sim$ 2.6 -- 4.2 arcmin. Their number distribution in the range of L $>$ 3 arcmin is fitted to a power-law with an exponent ($\alpha_{\rm L, non}$) of 3.7$\pm$0.2. While for filaments, their angular sizes distribute in a range of 3 -- 250 arcmin and mainly in $\sim$ 5.8 -- 12.1 arcmin. Their number distribution in the range of L $>$ 8 arcmin also is fitted to the power-law with an exponent of 1.9$\pm$0.1 ($\alpha_{\rm {L, fil}}$). According to the spiral structure of the Milky Way, we estimated the kinematical distances to sources using the Bayesian Distance Calculator in \cite{Reid2016}. We find that our molecular clouds in the Local arm may have a distance of $\sim$ 0.5 kpc, and those in the Perseus arm have a typical distance of $\sim$ 2 kpc. The molecular cloud with an angular size of 1 arcmin in the Local arm, has a physical scale of $\sim$ 0.15 pc, for that in the Perseus arm, its physical scale is $\sim$ 0.6 pc.

In Figure \ref{fig:fco}, we are based on the radial velocities of molecular clouds and roughly separate them into two groups, i.e., the near and far ranges.
As shown in Figure \ref{fig:fproj}, in the near range, the angular areas ($\mathcal{A}$) for non-filament structures are in a range of 2.5 -- 1450 arcmin$^{2}$ and concentrate on 6.5 -- 18.0 arcmin$^{2}$. Their number distribution in the range of $\mathcal{A}$ $>$ 10 arcmin$^{2}$ is fitted to a power-law with an exponent of 1.7$\pm$0.1 ($\alpha_{\rm A, non}$). Whereas, filaments distribute in a range of 8.75 -- 6.0$\times$10$^{4}$ arcmin$^{2}$ and mainly in 32 -- 147.4 arcmin$^{2}$, the exponent for their power-law distribution fitted in the range of $\mathcal{A}$ $>$ 100 arcmin$^{2}$ is 0.9$\pm$0.06 ($\alpha_{\rm A, fil}$). In the far range, the angular areas for non-filaments are in a range of 3 -- 180 arcmin$^{2}$ and focus on 6.75 -- 17.0 arcmin$^{2}$, their number distribution in the range of $\mathcal{A}$ $>$ 10 arcmin$^{2}$ is fitted to the power-law with an exponent of 2.0$\pm$0.1 ($\alpha_{\rm A, non}$). Filaments distribute in a range of 10 -- 2.1$\times$10$^{4}$ arcmin$^{2}$ and mainly in 36 -- 144.8 arcmin$^{2}$, the exponent for their power-law distribution fitted in the range of $\mathcal{A}$ $>$ 100 arcmin$^{2}$ is 1.0$\pm$0.1 ($\alpha_{\rm A, fil}$). The fitted exponent of the power-law for non-filaments is twice that for filaments, either in the near or far range, and the filaments tend to have a larger spatial size than non-filaments.

\subsection{$^{12}$CO emission intensity}
In Figure \ref{fig:fpara}, for the non-filaments, the peak intensities (I) of $^{12}$CO line emission are in a range of 1.4 -- 36.1 K and concentrate on 2.7 -- 4.2 K, and the number distribution in the range of I $>$ 3 K is fitted to a power-law with an exponent of 3.5$\pm$0.2 ($\alpha_{\rm I, non}$). Filaments distribute in a range of 2.0 -- 50.8 K and are mainly located in 4.0 -- 7.2 K, the fitted negative slope ($\alpha_{\rm I, fil}$) for the power-law distribution in the range of I $>$ 6 K is 2.8$\pm$0.07. 

Figure \ref{fig:fproj} presents the averaged $^{12}$CO velocity-integrated intensity, W$_{\rm ^{12}CO}$ = $\int$T$_{\rm mb}(l, b, v)dvdldb$/$\int$$dldb$, for molecular clouds. In the near range, for the non-filaments, the W$_{\rm ^{12}CO}$ ranges from 0.3 to 13.1 K km s$^{-1}$ and focuses on 0.8 -- 1.4 K km s$^{-1}$, the number distribution in the range of W$_{\rm ^{12}CO}$ $>$ 1.0 is fitted to a power law with an exponent of 3.9$\pm$0.3 ($\alpha_{\rm W, non}$). For filaments, the value of  W$_{\rm ^{12}CO}$ is in a range of 0.5 -- 12.3 K km s$^{-1}$ and concentrates on 1.2 -- 2.6 K km s$^{-1}$, the fitted power-law index for the number distribution in the W$_{\rm ^{12}CO}$ $>$ 2.0 is 3.1$\pm$0.2 ($\alpha_{\rm W, fil}$).
 
In the far range, for the non-filaments, the values of W$_{\rm ^{12}CO}$ are distributed in the range of 0.34 -- 15.9 K km s$^{-1}$ and mainly in 1.1 -- 2.0 K km s$^{-1}$, the exponent ($\alpha_{\rm W, non}$) for the fitted power-law distribution in the range of W$_{\rm ^{12}CO}$ $>$ 1.0 is 3.0$\pm$0.2. For the filaments, the values of W$_{\rm ^{12}CO}$ are in 0.6 -- 18.5 K km s$^{-1}$ and focus on 1.7 -- 3.8 K km s$^{-1}$, the fitted power-law index ($\alpha_{\rm W, fil}$) for the number distribution in the range of W$_{\rm ^{12}CO}$ $>$ 4.0 is 2.6$\pm$0.1. According to the analysis of beam filling factors ($\eta$) on molecular clouds using the MWISP CO data in \cite{Yanbeam2021}, the value of $\eta$ is about 0.5 for the non-filament with an angular size of $\sim$ 3 arcmin and about 0.8 for filaments in the range of 6 -- 12 arcmin. Taking the beam filling factor into account, we find that there are no large differences between the values of W$_{\rm ^{12}CO}$ for non-filaments and filaments, i.e., the H$_{2}$ column densities do not vary significantly among non-filaments and filaments.

Compared with the number distributions of angular size and W$_{\rm ^{12}CO}$ for non-filaments and filaments, we find the angular size that shows the most significant differences between non-filaments and filaments. The typical values of W$_{\rm ^{12}CO}$ for filaments are $\sim$1.5 times as large as that in non-filaments, if we take the beam filling factor of $\sim$ 0.5 into account for non-filaments and 0.8 for filaments, the W$_{\rm ^{12}CO}$ for non-filaments are more close to that for filaments. Their fitted slopes for the power-law distributions in the large-value ranges are also close. While the typical angular sizes for filaments are at least twice that for non-filaments, as well as their values of $\alpha_{\rm L}$.

\subsection{The integrated flux of $^{12}$CO(1-0) emission} 
Figure \ref{fig:fpara} presents the number distributions of the flux of $^{12}$CO(1-0) line emission for non-filaments and filaments, i.e., Flux = $\int$T$_{\rm mb}(p,p,v)dvdldb$. For the non-filaments, their $^{12}$CO(1-0) emission fluxes distribute in a range of 1.0 -- 2262 K km s$^{-1}$ arcmin$^{2}$ and concentrate on $\sim$ 6.7 -- 25.5 K km s$^{-1}$ arcmin$^{2}$, the number distribution in the range of Flux $>$ 30 K km s$^{-1}$ arcmin$^{2}$ is fitted to a power-law with an index of 1.7$\pm$0.1 ($\alpha_{\rm F, non}$). For the filaments, the values of Flux range from 5.6 to 3.9$\times$10$^{5}$ K km s$^{-1}$ arcmin$^{2}$ and mainly from 50.4 to 420 K km s$^{-1}$ arcmin$^{2}$, the fitted power-law index for the number distribution in the range of Flux $>$ 300 K km s$^{-1}$ arcmin$^{2}$ is 0.8$\pm$0.1 ($\alpha_{\rm F, fil}$). The filaments tend to have more $^{12}$CO(1-0) line emission flux than non-filaments. 

Furthermore, we calculated the $^{12}$CO(1-0) emission flux for the whole molecular clouds, the whole non-filaments, and the whole filaments in our catalog, respectively. In the $^{12}$CO(1-0) flux emission for the whole 18190 molecular clouds (3.7$\times$10$^{6}$ K km s$^{-1}$ arcmin$^{2}$), $\sim$ 90$\%$ are from filaments (3.4$\times$10$^{6}$ K km s$^{-1}$ arcmin$^{2}$), $\sim$ 9$\%$ are from non-filaments (3.3$\times$10$^{5}$ K km s$^{-1}$ arcmin$^{2}$ ), and the rest $\sim$ 1$\%$ are from unresolved sources. Although non-filaments are dominant in the number of discrete molecular clouds, filaments are the main contributor of the $^{12}$CO emission flux.

\subsection{Velocity span}

Figure \ref{fig:fpara} presents the number distributions of the velocity spans (V$_{\rm span}$) for non-filaments and filaments. For non-filaments, the values of V$_{\rm span}$ range from 0.67 to 22.4 km s$^{-1}$ and concentrate on $\sim$ 1.7 -- 3.0 km s$^{-1}$, the number distribution in the range of V$_{\rm span}$ $>$ 3 km s$^{-1}$ is fitted to a power-law function with an exponent of 4.5$\pm$0.2 ($\alpha_{\rm V, non}$). For the filaments, the V$_{\rm span}$ is distributed in a wide range of 1.0 -- 43.4 km s$^{-1}$, and the values for most sources are located in a range of 3.2 -- 6.5 km s$^{-1}$. The fitted power-law index for the number distribution in the range of V$_{\rm span}$ $>$ 6 km s$^{-1}$ is 3.1$\pm$0.2. Compared with non-filaments, the filaments with larger angular sizes, have wider velocity spans.

\section{Discussion}
\subsection{Human biases}
Since the morphological classification based on visual inspections are conducted independently by three authors, there may exist human biases. As mentioned in Section 3.2, the agreement rates of morphological classification by two individuals are about 60 $\%$ -- 80 $\%$. The agreement level of the classified results of three persons is about 60 $\%$.

For unresolved sources and resolved clouds, we divide them by inspecting the CO emission area for clouds using the criterion of 5 pixels $\times$ 5 pixels. These unresolved morphologies with pixels less than 5 $\times$ 5 are clear and easy to confirm, which are less affected by human biases. However, the categories of non-filaments and filaments are mainly determined by the estimation of aspect ratios of filamentary segments. An elongated and narrow structure with its aspect ratio $\gtrsim$ 4 is a filament, the network connected by a set of multiple filamentary segments also is the filamentary cloud. Networks usually present large spatial sizes, their hierarchical and filamentary structures are easy to experientially validate by eyes.

The classification of the extended structures (non-filaments in Figure 15) and a single filament (filaments in Figure 12) may be affected by the visual estimation of their aspect ratios. According to the distribution of the angular size of clouds in Figure \ref{fig:fpara}, the non-filaments with angular size larger than $\sim$ 8 arcmin, which is the typical angular size for filaments, take a percentage of $\sim$ 2 $\%$ in the total molecular clouds. Their morphologies may be more easily affected by human biases. 
%Thus there might be about 2 $\%$ errors caused by human bias in the morphological fractions of filaments and non-filaments, respectively.

In addition, machine learning techniques, which are more objective than visual inspections, have been widely used in the morphological classification for galaxies \citep{Banerji2010, Ghosh2020, Cheng2020, Walmsley2020}, the detection of interstellar shells, bubbles \citep{Beaumont2014, VanOotr2019}, and molecular outflows \citep{Zhang2020, Xu2020}. The agreement rates between galaxy classifications made by humans and machine learning approaches are over 90 $\%$ \citep{Beck2018, Cheng2020, Cheng2021, Walmsley2021}. For the morphological classifications of compact star clusters and associations, the level of agreement is about 82 $\%$ \citep{Whitmore2021}.

\subsection{Observational effect}
\subsubsection{Finite angular resolution}
Due to the limited angular resolution of $^{12}$CO line, the resolved morphologies of molecular clouds may be influenced by their distances. As mentioned in Section 4.2, the physical resolutions are different for molecular clouds with different distances.

In Figure \ref{fig:fco}, in the radial velocity range of ($-$95, 25) km s$^{-1}$, the 18190 molecular clouds are grouped into two parts, corresponding to the near and far ranges, respectively. Molecular clouds in the near range are in the velocity range of ($-$30, 25) km s$^{-1}$, clouds in the far range are in ($-$95, $-$30) km s$^{-1}$. In a total of 18190 molecular clouds, there are 9544 clouds in the near range, including 1602 (16.8 $\%$) unresolved sources, 6747 (70.7 $\%$) non-filaments and 1195 (12.5 $\%$) filaments. While, 8646 clouds located in the far range, contain 2846 (32.9 $\%$) unresolved sources, 4933 (57.1 $\%$) non-filaments and 867 (10.0 $\%$) filaments. 

\begin{figure*}
\plotone{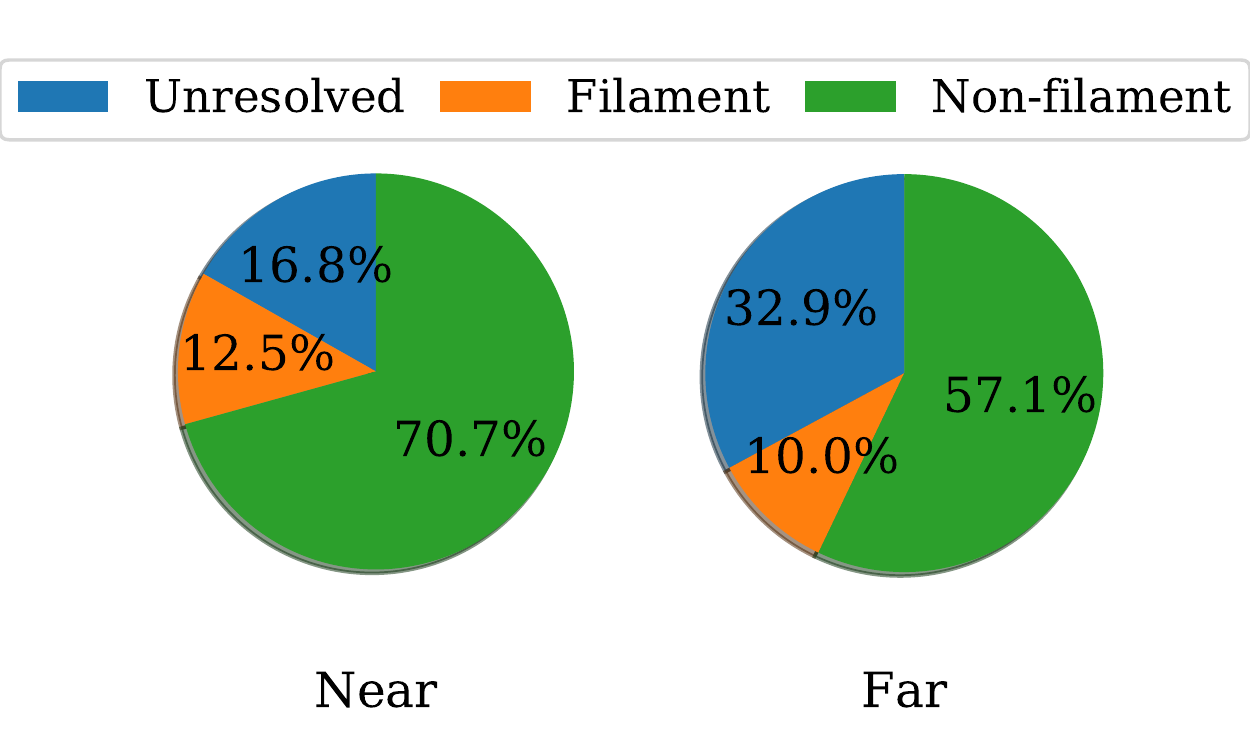}
\caption{The morphological fractions of molecular clouds in the near and far ranges, respectively.}
\label{fig:ffrac}
\end{figure*}

Figure \ref{fig:ffrac} shows the morphological fractions of molecular clouds, i.e., the fractions of unresolved sources, non-filaments, and filaments in the near and far ranges, respectively. The fraction of unresolved sources in the far range is nearly twice that in the near range. Due to the limited angular resolution, molecular clouds in the far range with larger distances usually have smaller angular sizes. There are about $\sim$ 3$\%$ difference in the filament fractions of molecular clouds in the near and far ranges, which is not very significant, as a result of that most of filaments have large spatial sizes. 

It should be noted that the physical resolutions of our observed molecular clouds are still finite. The non-filaments may present the filamentary and hierarchical structures under higher spatial resolutions.

\subsubsection{Beam dilution effect}
The beam smoothing on the observations of molecular clouds, can diminish their brightness temperatures and flatten their structures. That may affect the observed morphologies of molecular clouds, particularly for the molecular clouds with small angular sizes. In this work, the resolved molecular clouds have the observed angular areas larger than $\sim$ 2.5 arcmin $\times$ 2.5 arcmins, corresponding to a diameter of $\sim$ 3 beam size. We estimate the actual angular size (L$_{\rm act}$) of molecular clouds as $\sqrt{\rm L_{\rm obs}^{2} - \Theta^{2}}$, where the L$_{\rm obs}$ is the observed angular size, $\Theta$ is the beam size. The value of L$_{\rm act}$ is about 2.3 arcmin for the molecular clouds with the L$_{\rm obs}$ of 2.5 arcmin. Thus the effect of beam dilution on the angular sizes of resolved clouds is less than $\sim$ 10$\%$. These errors of the estimated aspect ratios of structures, which are caused by the beam dilution effect, should be less than that caused by human bias.

In addition, as mentioned in Section 4.3, the observed values about the W$_{\rm ^{12}CO}$ for filaments and non-filaments have small differences. That also supports that the beam dilution effect should not be severe for non-filaments. This implies that there is only a small portion of non-filaments that are actually slim filaments but mis-classified as non-filaments, due to their radial widths being broadened by beam convolution.

It should be noted that the finite angular resolution and sensitivity will limit the angular sizes of the observed structures. Maybe there are more tiny and hierarchical structures, which are not detected and resolved by PMO 13.7-m telescope. The follow-up observations with higher resolution, can help us to reveal the structures of the unresolved sources. In this work, we mainly discuss the morphological classification of resolved clouds in our catalog.

\subsubsection{Projectional effect}
The morphologies of molecular clouds are determined by their projected shapes along the line of sight, thus the morphologies of clouds may be influenced by the projection effect. For a filament, we assume the angle between its long axis and the line of sight is $\theta$, thus its aspect ratio ($A_{\rm o}$) becomes $A_{\rm o}$$\rm \sin$$\theta$. That may cause a portion of filaments ($A_{\rm o}$ $\gtrsim$ 4) to present the morphologies of non-filaments ($A_{\rm o}$ $\lesssim$ 4). While the line-of-sight projection changes the aspect ratios of molecular clouds, as well as their H$_{2}$ column densities (N$_{\rm H_{2}}$). If the aspect ratios of filaments become $A_{\rm o}$$\rm \sin$$\theta$, their H$_{2}$ column densities become N$_{\rm H_{2}}$/$\rm \sin$$\theta$ as well.

Figure \ref{fig:fproj} shows the number distributions of the averaged velocity-integrated $^{12}$CO line intensity, i.e., W$_{\rm ^{12}CO}$ = $\int$T$_{\rm mb}(l,b,v)dldbdv$/$\int$$dldb$, and the angular areas for molecular clouds. Furthermore, We could estimate the H$_{2}$ column density (N$_{\rm H_{2}}$) using the N$_{\rm H_{2}}$ = X$_{\rm CO}$W$_{\rm ^{12}CO}$, where X$_{\rm CO}$ = 2$\times$10$^{20}$ cm$^{-2}$ (K km s$^r{-1}$)$^{-1}$ is the CO-to-H$_{2}$ conversion factor \citep{Bolatto2013}. In the near range, the values of W$_{\rm ^{12}CO}$ for non-filaments mainly distribute in a range of 0.8 - 1.4 K km s$^{-1}$ (N$_{\rm H_{2}}$ $\sim$ 1.6 $\times$10$^{20}$ -- 2.8 $\times$10$^{20}$ cm$^{-2}$), those values for filaments concentrate on a range of 1.2 -- 2.6 K km s$^{-1}$ (N$_{\rm H_{2}}$ $\sim$ 2.4 $\times$10$^{20}$ -- 5.2 $\times$10$^{20}$ cm$^{-2}$). In the far region, the values of W$_{\rm ^{12}CO}$ for non-filaments mainly locate in (1.1, 2.0) K km s$^{-1}$ (N$_{\rm H_{2}}$ $\sim$ 2.2 $\times$10$^{20}$ -- 4.0 $\times$10$^{20}$ cm$^{-2}$), the typical range of W$_{\rm ^{12}CO}$ for filaments is from 1.7 to 3.8 K km s$^{-1}$ (N$_{\rm H_{2}}$ $\sim$ 3.4 $\times$10$^{20}$ -- 7.6 $\times$10$^{20}$ cm$^{-2}$).

Due to the projection effect, a portion of filaments may present non-filaments. Meanwhile, their H$_{2}$ column densities could be N$_{\rm H_{2}, fil}$/$\sin$$\theta$, where N$_{\rm H_{2}, fil}$ is the H$_{2}$ column density of the unprojected filament. Filaments have $A_{o}$ $\gtrsim$ 4, if $\sin$$\theta$ $\lesssim$ 0.5, the value of $A_{o}$$\sin$$\theta$ tends to be smaller than 4, thus the filaments may be projected as non-filaments. Meantime, the H$_{2}$ column densities for these projected non-filaments, may be larger than N$_{\rm H_{2}, fil}$/$\sin$($\theta$) $\sim$ 2N$_{\rm H_{2}, fil}$. Thus, the non-filaments, which have N$_{\rm H_{2}}$ larger than 2N$_{\rm H_{2}, fil}$, may be affected by the projection. As the W$_{\rm ^{12}CO}$ for filaments mainly focus on 1.5 -- 3.5 K km s$^{-1}$, thus the non-filaments with W$_{\rm ^{12}CO}$ larger than $\sim$ 5 K km s$^{-1}$ tend to be projected by filaments. Thus $\sim$ 1$\%$ molecular clouds in our sample classified as non-filaments, probably are due to the line-of-sight projection of filaments. If we take the beam filling factor of 0.5 into account for non-filaments, the fraction is up to $\sim$ 3$\%$.

We note that the W($^{12}$CO) for molecular clouds are distributed in a range of more than an order of magnitude, intrinsically. Thus the values of W$_{\rm ^{12}CO}$, even larger than $\sim$ 5 K km s$^{-1}$, may also be the original values of non-filaments, not due to the enhancement of the column density caused by the projection of filaments.

Next, we attempt to discuss the projection effect of molecular clouds, geometrically. We assume the actual 3-D structure of a filament is a cylinder with a diameter of $D$ and a height of $L$. The $D$ will correspond to the width of the filament, and $L$ is the length. Furthermore, we assume the variable of $D$ is normally distributed,
\begin{equation}
f(D \vert \mu, \sigma^2) = \frac{1}{\sigma \sqrt{2\pi}} e^{-\frac{1}{2}\left(\frac{D - \mu}{\sigma} \right)^{2}},
\end{equation}
where $\mu$ is the mean of the $D$, $\sigma$ is its standard deviation. 

$L$ is sampled from a power-law distribution
\begin{equation}
p(L) = kL^{-\alpha},
\end{equation}
with a $< L <$ b . From the requirement that $\int_{a}^{b}p(L)dL = 1$, we have
\begin{equation}
k = \frac{1-\alpha}{b^{1-\alpha} - a^{1- \alpha}} .
\end{equation}
For the special case $\alpha$ =1,
\begin{equation}
k= \frac{1}{ln(b/a)} .
\end{equation}

For a filament with a diameter of $D$ and a height of $L$, its aspect ratio is $L/D$ and its area is $LD$. $\theta$ is the angle between the line of sight and the long axis of a filament. If the filament has an angle $\theta$ with the line-of-sight direction, when viewed from the radial direction, its projected aspect ratio will be approximately 
\begin{equation}
\frac{L \rm sin\theta}{D},
\end{equation}
and the projected area will be approximately
\begin{equation}
DL \rm sin\theta .
\end{equation}

According to the distribution of the angular sizes of resolved molecular clouds in Figure \ref{fig:fpara}, their angular sizes are about larger than 2.5 arcmin, and the fitted power-law exponent is about 2.0. After visually inspecting the cataloged filaments, the largest length of a filament is about 120 arcmin. Thus we assume $\alpha$=2.0, a = 2.5, and b=120 in the power-law distribution of $L$. In the normal distribution of $D$, we assume $\mu$ = 2.5, which is the minimal resolved scale, and $\sigma$=1.

\begin{figure*}
\plotone{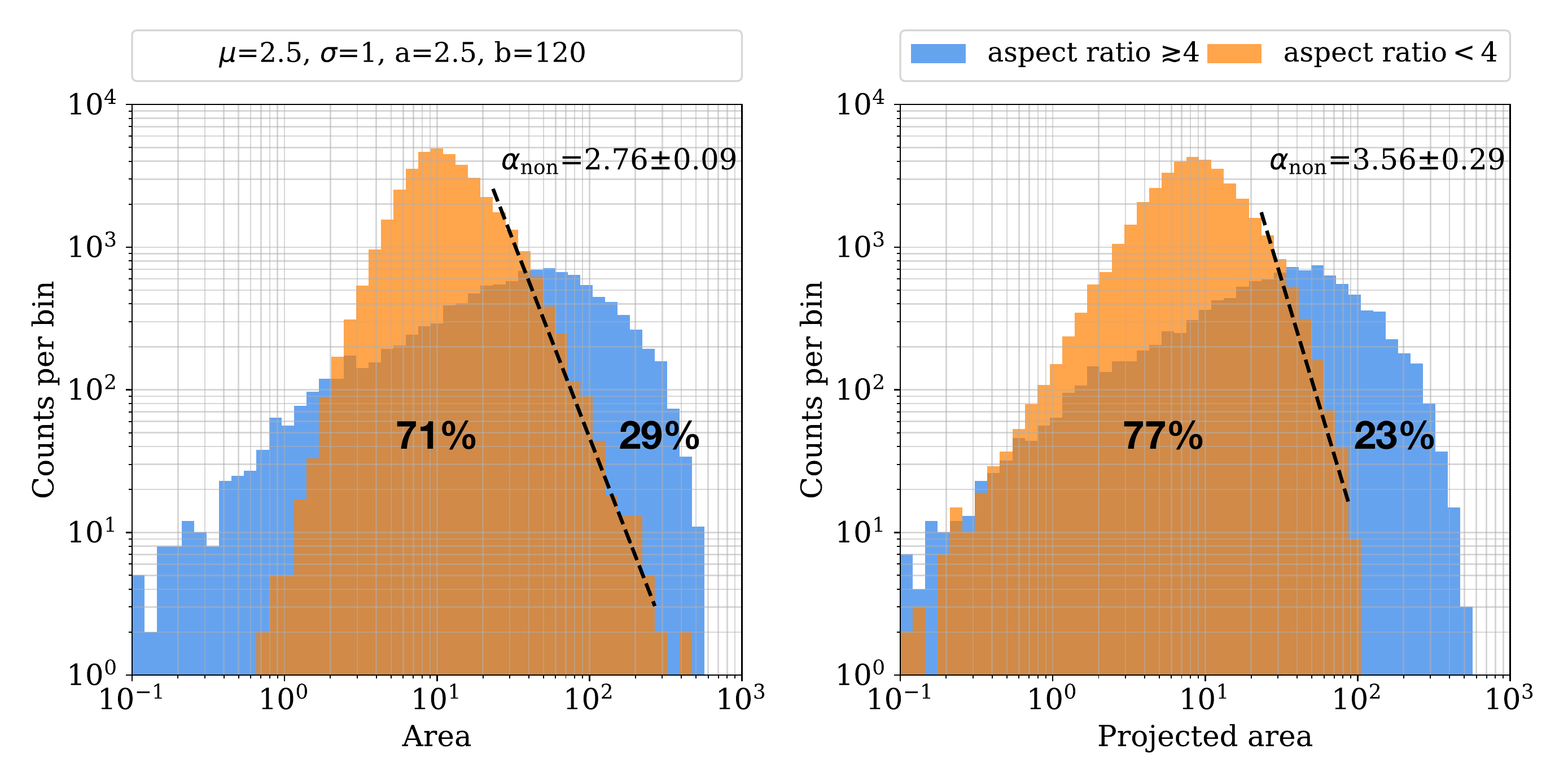}
\caption{\textbf{Left panel}: the number distribution of the area ($LD$) for the simulated clouds. \textbf{Right panel}: the number distribution of the projected area ($LD$sin$\theta$) for the simulated clouds. The dashed-black lines fit their power-law distributions, dN/dX $\propto$ X$^{-(\alpha_{\rm X}+1)}$, whose negative slopes are noted in each panel.}
\label{fig:fsim}
\end{figure*}

In Figure \ref{fig:fsim}, we choose two subsamples based on their observed aspect ratios ($A_{o}$ = $L$ sin$\theta$/$D$), i.e., non-filaments ($A_{o}$ $<$ 4) and filaments ($A_{o}$ $\gtrsim$ 4), and present the number distributions of their areas ($LD$) and projected areas ($LD$sin$\theta$), respectively. We also estimated the number fractions of non-filaments and filaments in the samples, as well as those values in the samples after the projection. We find that the number fraction of filaments is 29$\%$ in the samples, and changes about 6$\%$ after the projection. In addition, we fit the power-law slopes for the number distribution of areas of non-filaments. The fitted exponents for non-filaments change from 2.76$\pm$0.09 to 3.56$\pm$0.29, after the projection. That is due to the non-filaments with larger angular areas usually have aspect ratios being close to 4, their number distribution is easily changed by the line-of-sight projection of filaments.

In addition, we find the distribution of sampled filaments in the range with areas larger than 10$^{2}$ is different from that for our observed filaments. We should note that the assumptions of cylinders and their parameters are extremely simplified. A portion of the observed filamentary clouds, especially for that with larger areas, contain more than one filament segment. This sample just is a side of reflection of the projection effect on the morphologies of molecular clouds.

\subsection{Implications for the molecular clouds evolution in the Milky Way}
It is interesting to explore the possible relation between filaments and non-filaments. The systematic difference between them is the angular size, and their average H$_{2}$ column densities do not significantly vary among them. Previous studies suggest that the non-filaments are formed by the fragmentation of filaments, which is caused by the gravitational instability \citep{Larson1985, Nakamura1993, Bonnell1992, Jackson2010, Clarke2017, Clarke2020}, or the agglomeration of non-filaments build up filaments \citep{Dobbs2008, Tasker2009, Dobbs2011, Smith2016, Baba2017, Kobayashi2017}. It is important to provide observational evidence to confirm these mechanisms systematically. Further observations and analysis, such as the dense gas fractions in non-filaments and filaments, may help us to further understand these questions. This visual classification also is a starting point for future attempts to improve automated morphological classification for a giant sample of molecular clouds from the whole MWISP CO project using machine learning techniques.

Furthermore, our observational results can be compared with the results of numerical simulations on the evolution of molecular clouds in Milky Way-like galaxies, such as the formation and physical properties of larger-scale filaments \citep{Smith2020, Zucker2019, Hoemann2021, Burkert2004} and the time evolution of molecular clouds \citep{Jeffreson2021}.

\section{Conclusions}

We use an unbiased catalog of 18190 molecular clouds, which is extracted from the $^{12}$CO(1-0) data of MWISP survey in the Galactic plane over 104.75$^{\circ}$ $< l <$ 150.25$^{\circ}$, $-$5.25$^{\circ}$ $< b <$ 5.25$^{\circ}$, and $-$25 km s$^{-1}$ $<$ $V_{\rm LSR}$ $<$ 95 km s$^{-1}$, to classify the morphology of molecular clouds as unresolved sources, non-filaments, and filaments. We further statistically analyze the parameters for non-filaments and filaments. The main conclusions are:

\begin{enumerate}

\item The 18190 molecular clouds are visually divided into unresolved and resolved ones, and the resolved ones are further classified as non-filaments and filaments. In the 18190 molecular clouds, $\sim$ 25$\%$ are unresolved sources, $\sim$ 64$\%$ are non-filaments, and $\sim$ 11$\%$ are filaments.

\item We calculate the flux of $^{12}$CO(1-0) line emission of the whole 18190 molecular clouds. Filaments contribute $\sim$ 90$\%$ to the total flux of the whole sample, while non-filaments contribute $\sim$ 9$\%$, and the rest 1$\%$ is from unresolved sources. 

\item We present the number distributions of parameters for the whole sample, including the angular size, velocity span, peak intensity, and flux of $^{12}$CO(1-0) line emission. Their distributions in the large-value ranges can be fitted by a power-law. The fitted exponent for the power-law distribution of angular sizes is $\sim$ 1 for filaments, and 2 for non-filaments. After comparing the physical parameters of filaments with that of non-filaments, the major difference between filaments and non-filaments lies in their spatial scales. The filaments tend to have larger spatial scales. The H$_{2}$ column density does not vary significantly among them.  

\end{enumerate}

\begin{acknowledgements}
We gratefully thank the anonymous referee whose constructive review report helped improve the quality of this paper. This research made use of the data from the Milky Way Imaging Scroll Painting (MWISP) project, which is a multi-line survey in $^{12}$CO/$^{13}$CO/C$^{18}$O along the northern galactic plane with PMO-13.7m telescope. We are grateful to all the members of the MWISP working group, particularly the staff members at PMO-13.7m telescope, for their long-term support. MWISP was sponsored by National Key R\&D Program of China with grant 2017YFA0402701 and CAS Key Research Program of Frontier Sciences with grant QYZDJ-SSW-SLH047. JY is supported by National Natural Science Foundation of China through grant 12041305. FD is supported by the National Natural Science Foundation of China through grant 11873094.

\end{acknowledgements}

\bibliographystyle{aasjournal}
\bibliography{Morphology.bib}

\appendix

%\section{Appendix information}

\begin{figure*}[h!]
\plotone{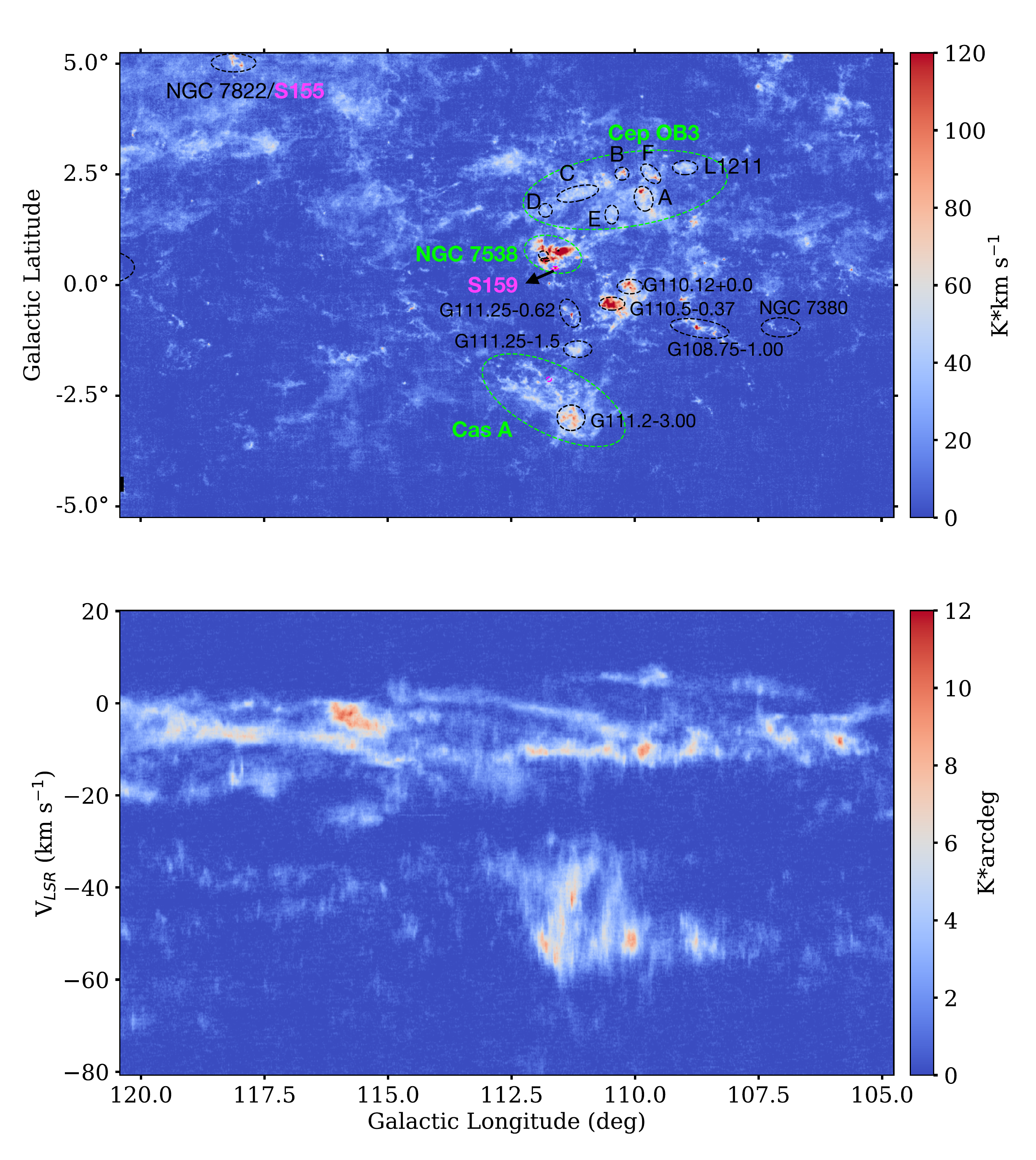}
\caption{\textbf{Top panel}: the velocity-integrated intensity map of $^{12}$CO(1-0) emission in the region with 104.75$^{\circ}$ $< l <$ 120$^{\circ}$ and $|b| <$ 5.25$^{\circ}$. The integrated velocity ($V_{\rm LSR}$) ranges from $-$80 km s$^{-1}$ to 20 km s$^{-1}$. The coordinates of clumps (A -- E) in Cepheus OB3 are from \citet{Yu1996}.
The HII region NGC 7538 that belongs to the Cas OB2 is refereed in \citet{Fallscheer2013}. The Molecular clouds near Cassiopeia A and NGC 7538 are from \citet{Ungerechts2000}. 
The Galactic young star cluster NGC 7380 is refereed as \citet{Chen2011}.
The HII region S155 and its neighboring molecular cloud NGC 7822 are refereed in \citet{Yang1992}. \textbf{Bottom panel}: the latitude-integrated intensity map of $^{12}$CO(1-0) emission in the region with 104.75$^{\circ}$ $< l <$ 120$^{\circ}$ and $-$80 km s$^{-1}$ $< V_{\rm LSR} <$ 20 km s$^{-1}$, the integrated latitude $b$ ranges from $-$5.25$^{\circ}$ to 5.25$^{\circ}$.}
\label{fig:fco_partI}
\end{figure*}

\begin{figure*}[h!]
\plotone{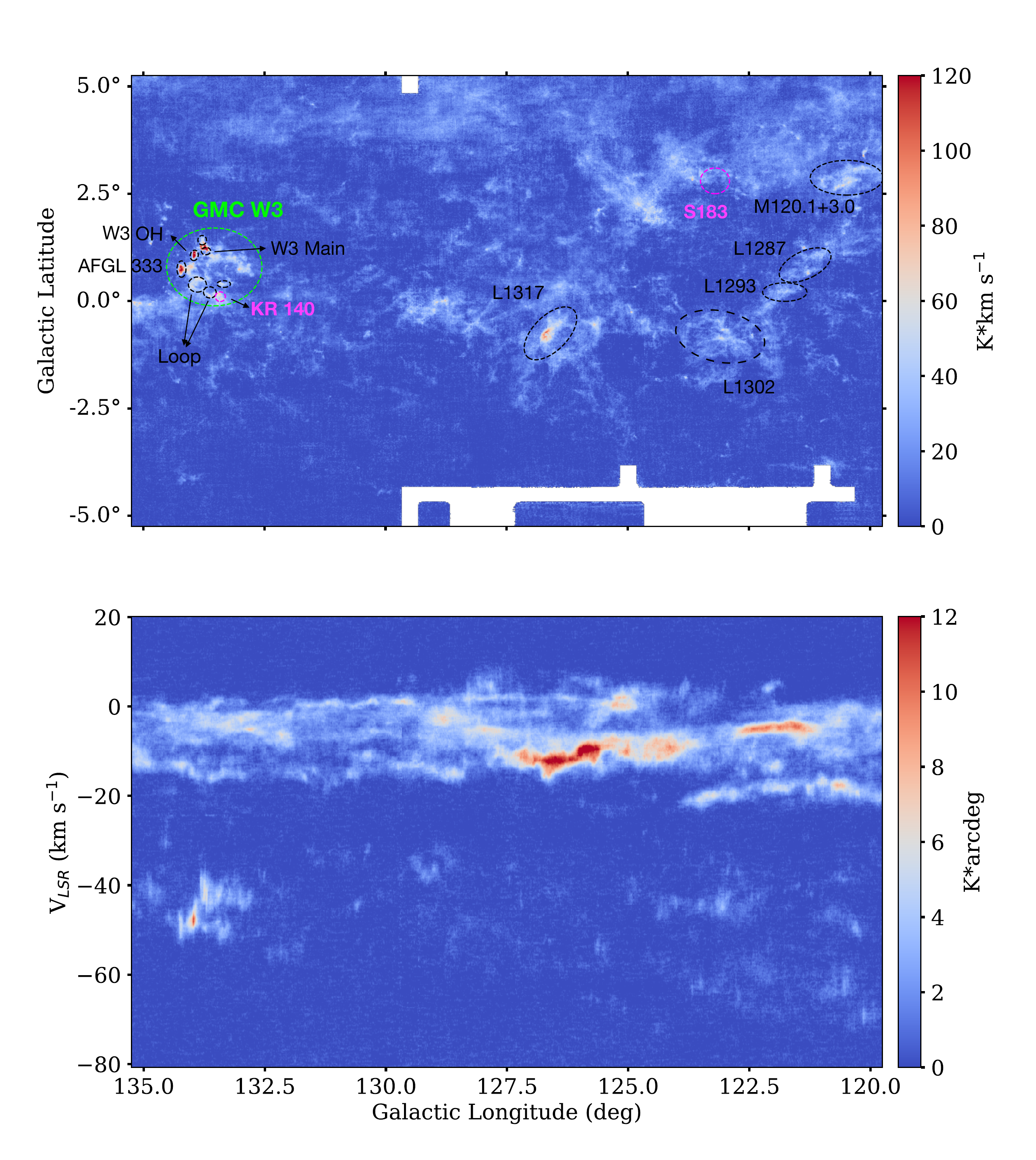}
\caption{\textbf{Top panel}: the velocity-integrated intensity map of $^{12}$CO(1-0) emission in the region with 120$^{\circ}$  $< l <$ 135$^{\circ}$ and $|b|$ $<$ 5.25$^{\circ}$. The integrated velocity ($V_{\rm LSR}$) range is (-80 20) km s$^{-1}$. The molecular cloud M120.1+3.0 in Cepheus OB4 is from \citet{Yang1990}.
The molecular cloud L1287 is from \citet{Yang1991}.
The molecular cloud L1293 is refereed as \citet{Yang1990}.
The HII region S187 and its environment L1317 are from \citet{Joncas1992}.
The HII region S183 is refereed in \citet{Landecker1992}.
The GMC W3 is refereed in \citet{RiveraIngraham2015}. \textbf{Bottom panel}: the latitude-integrated intensity map of $^{12}$CO(1-0) emission in the region with 120$^{\circ}$ $< l <$ 135$^{\circ}$ and $-$80 km s$^{-1}$ $< V_{\rm LSR} <$ 20 km s$^{-1}$, the integrated latitude $b$ ranges from $-$5.25$^{\circ}$ to 5.25$^{\circ}$.}
\label{fig:fco_partII}
\end{figure*}

\begin{figure*}[h!]
\plotone{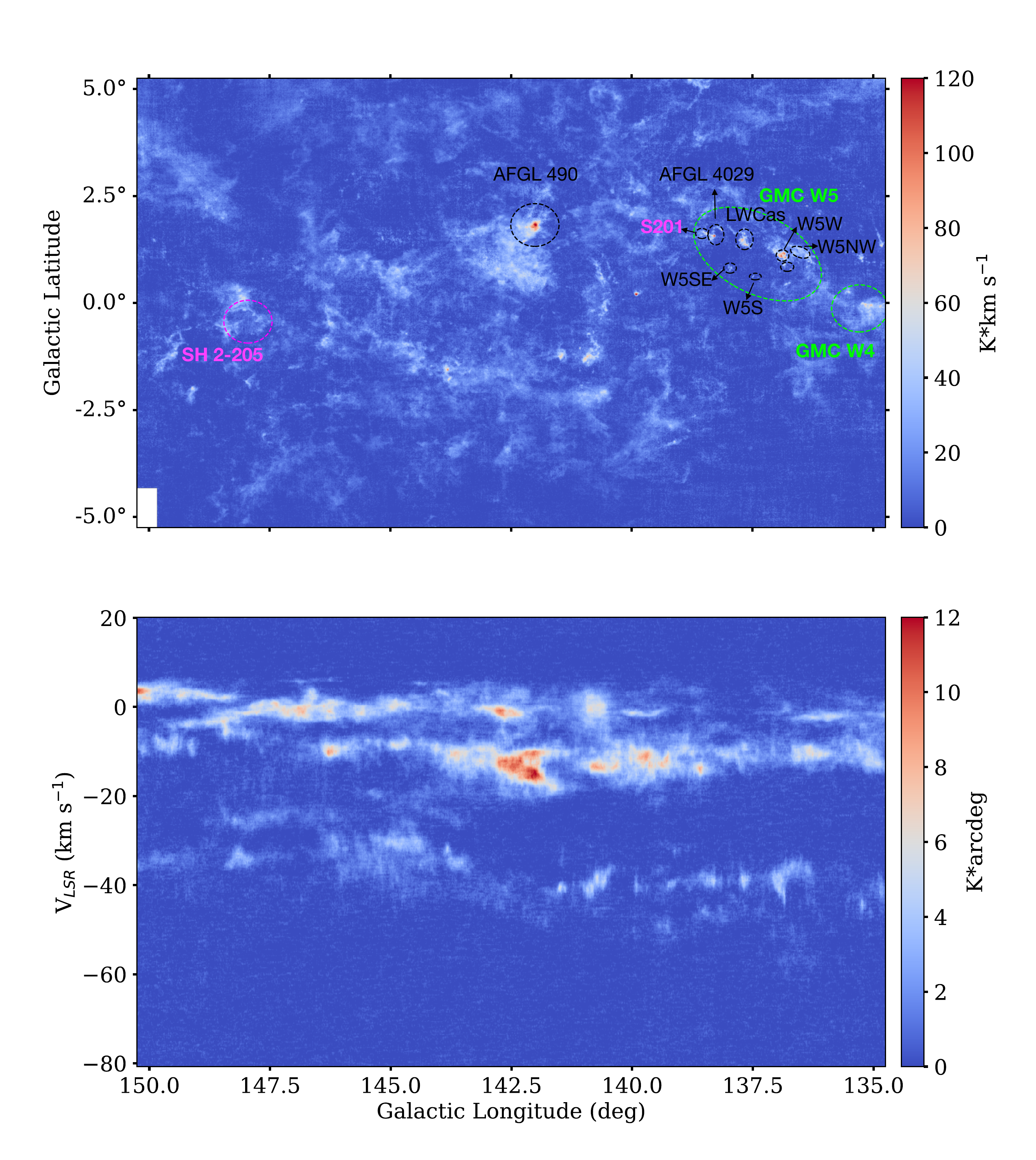}
\caption{\textbf{Top panel}: The velocity-integrated intensity map of $^{12}$CO(1-0) emission in the region with 135$^{\circ}$ $< l <$ 150.25$^{\circ}$ and $|b|$ $<$ 5.25$^{\circ}$. The integrated velocity ($V_{\rm LSR}$) ranges from $-$80 km s$^{-1}$ to 20 km s$^{-1}$. The GMC W3, W4 and W5 is refereed as \citet{RiveraIngraham2015, Bieging2011, Ginsburg2011}, respectively. \textbf{Bottom panel}: the latitude-integrated intensity map of $^{12}$CO(1-0) emission in the region with 135$^{\circ}$ $< l <$ 150.25$^{\circ}$ and $-$80 km s$^{-1}$ $< V_{\rm LSR} <$ 20 km s$^{-1}$, the integrated latitude $b$ ranges from $-$5.25$^{\circ}$ to 5.25$^{\circ}$.}
\label{fig:fco_partIII}
\end{figure*}

\begin{figure*}
\plotone{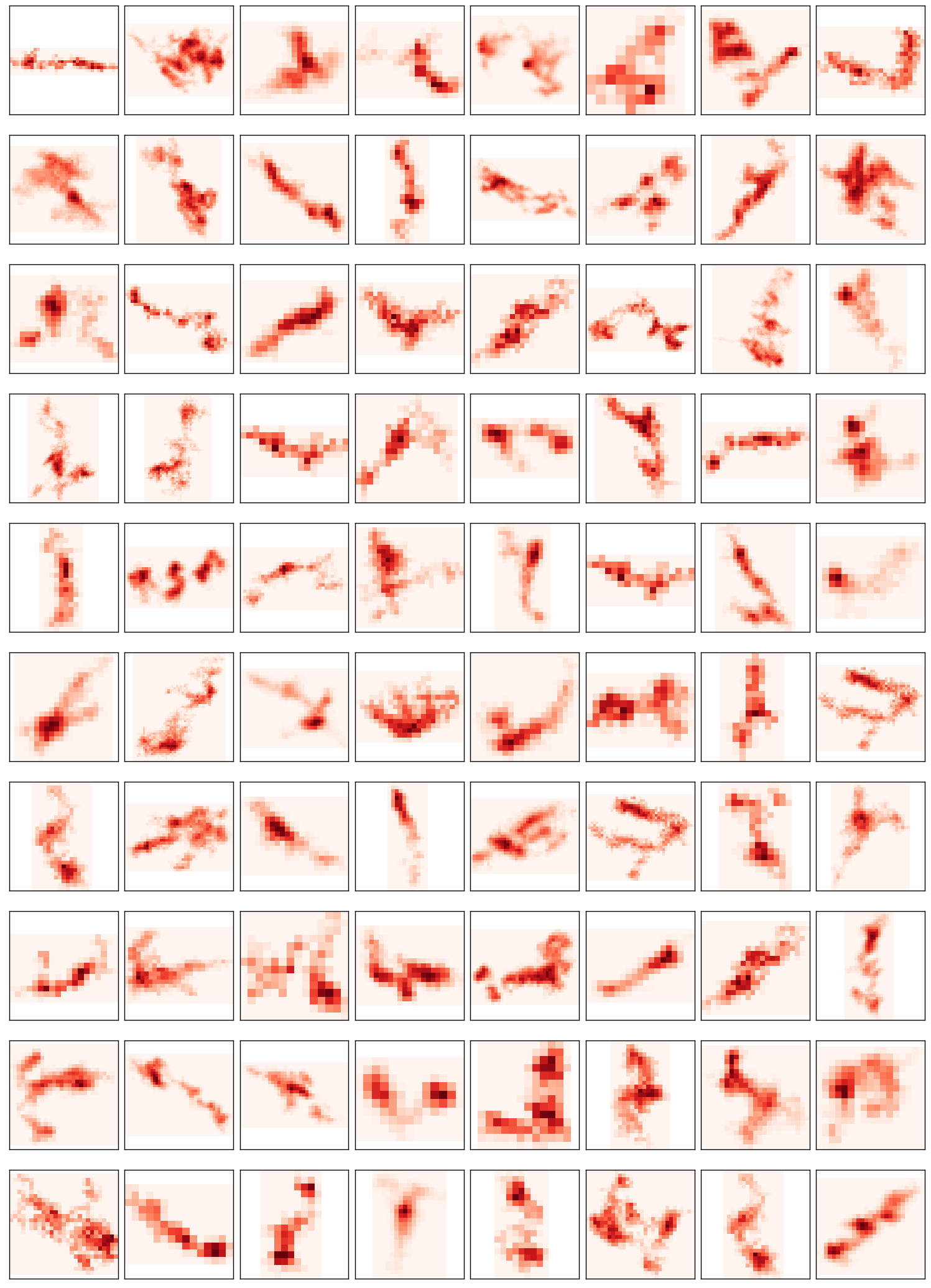}
\caption{Example images of filaments, which are randomly selected from molecular clouds presenting the elongated and narrow structures with $A_{o}$ $\gtrsim$ 4.} 
\label{ffilament}
\end{figure*}

\begin{figure*}
\plotone{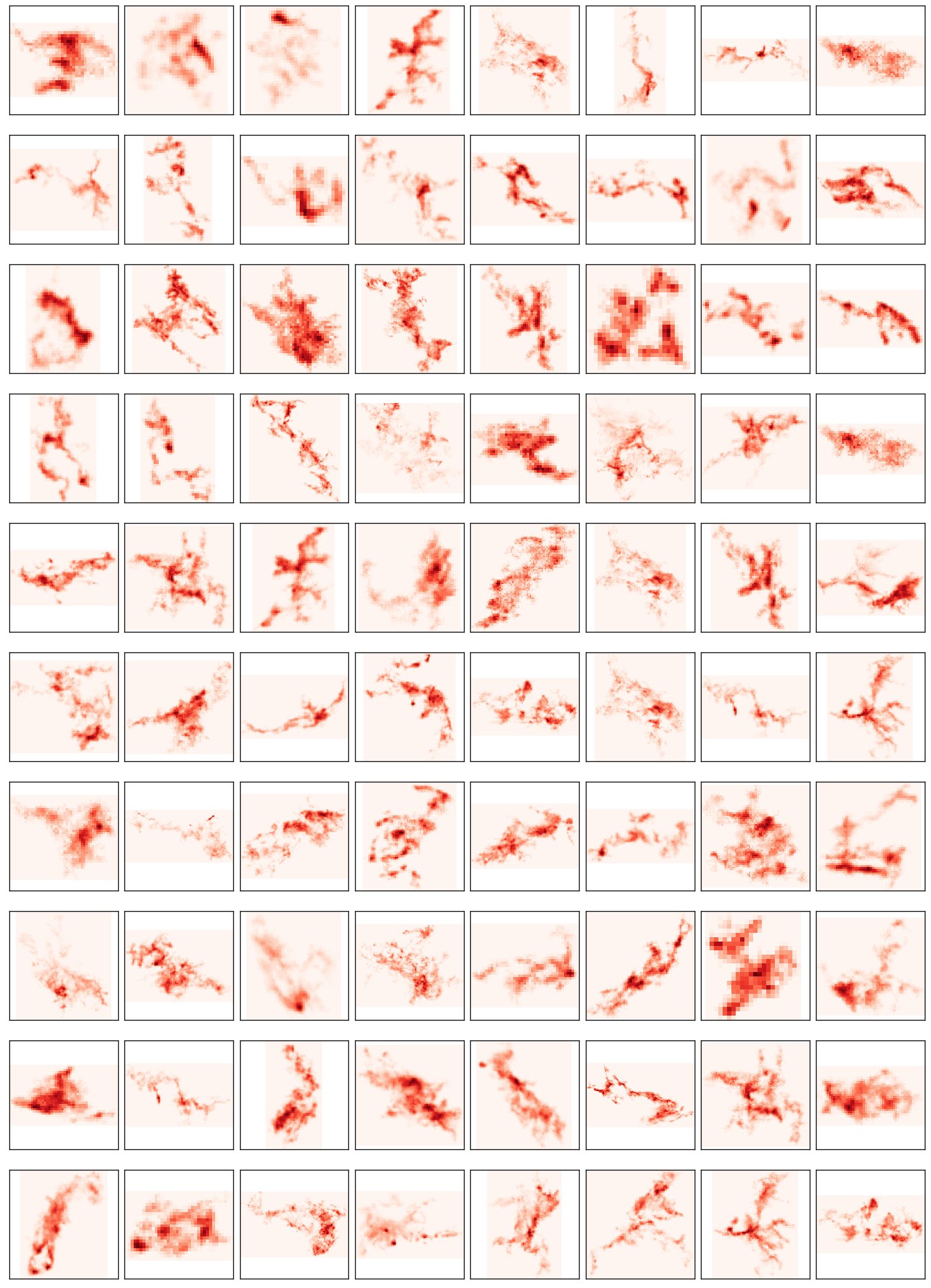}
\caption{Example images of filaments, which are randomly selected from molecular clouds built by a system of filaments.} 
\label{fnetwork}
\end{figure*}

\begin{figure*}[th!]
\plotone{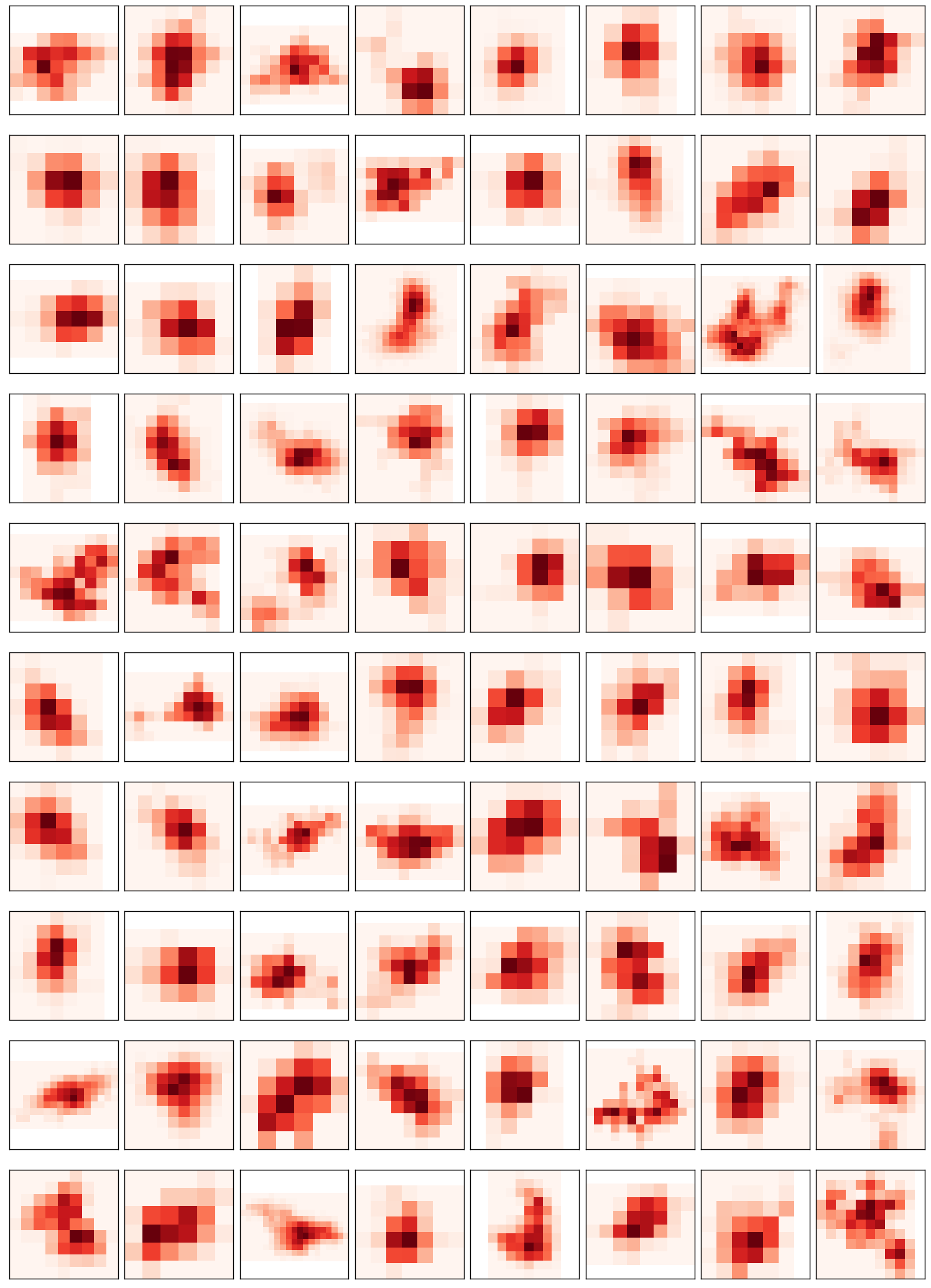}
\caption{Example images of non-filaments , which are randomly selected from molecular clouds showing clumpy structures.} 
\label{fclump}
\end{figure*}

\begin{figure*}
\plotone{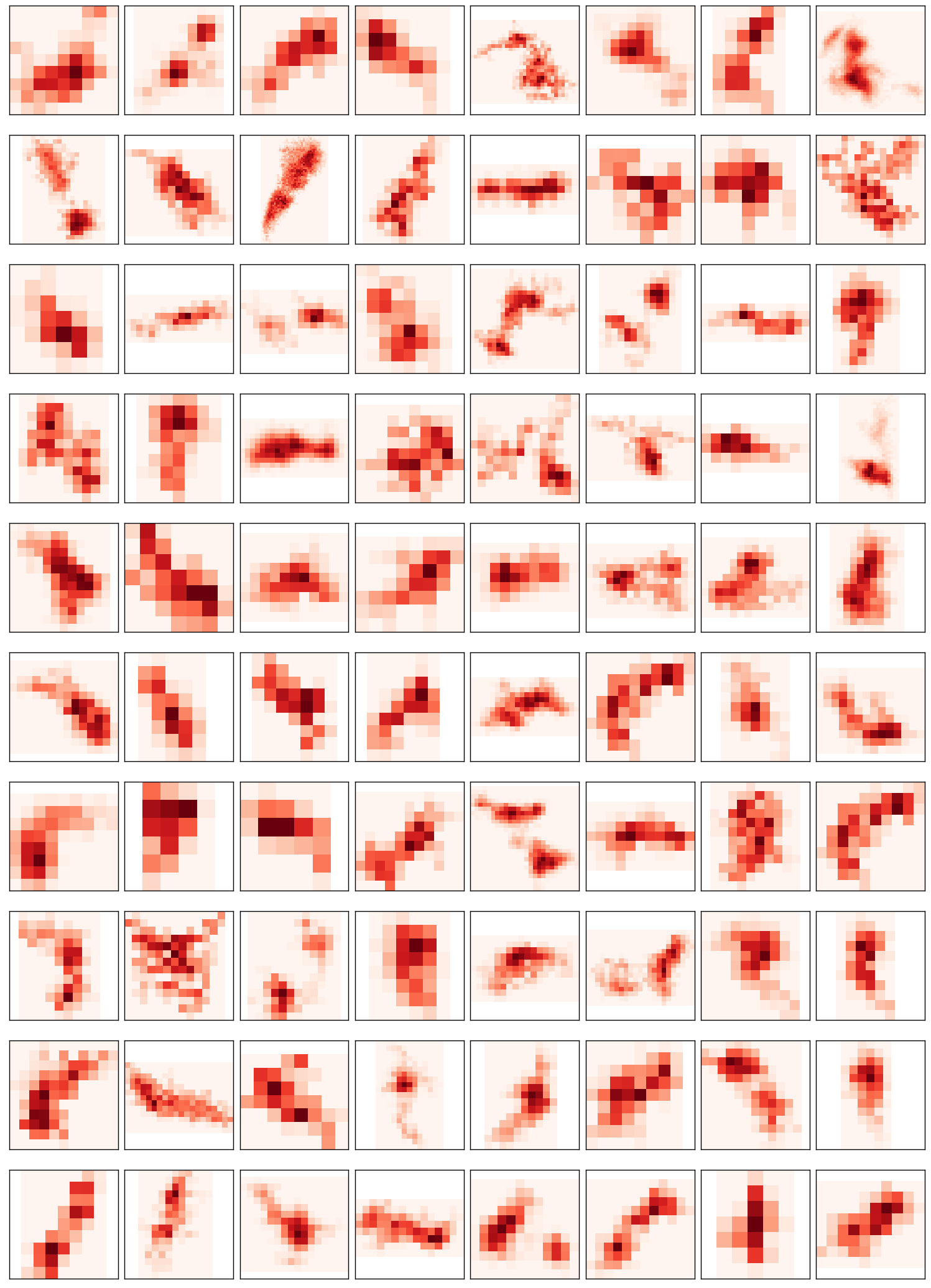}
\caption{Example images of Non-filaments, which are randomly selected from molecular clouds exhibiting extended structures with $A_{o}$ $\lesssim$ 4.} 
\label{fes}
\end{figure*}


%\bibliography{sample63}{morph_ref.bib}



\end{document}